\documentstyle[aas2pp4]{article}

\received{July 24 1999}
\accepted{23 September 1999}
\journalid{530}{February 10 2000}

\slugcomment{ApJ in press}

\lefthead{de Mello et al.}
\righthead{B stars as a Diagnostic of Star-formation at Low and High Redshift}

\def\ltsim{\lower.5ex\hbox{$\; \buildrel < \over \sim \;$}}
\begin{document}

\title{B stars as a diagnostic of star-formation at low and high redshift}

\author{Du\'{\i}lia F. de Mello}       
\affil{Space Telescope Science Institute\footnote{Operated
 by AURA for NASA under contract NAS5-26555}, 3700 San Martin Drive, Baltimore,
       MD 21218\\
       e-mail: demello@stsci.edu}
       
\author{Claus Leitherer}
\affil{Space Telescope Science Institute, 3700 San Martin Drive, Baltimore,
       MD 21218\\
       e-mail: leitherer@stsci.edu}

\and

\author{Timothy M. Heckman}
\affil{Physics and Astronomy Department, Johns Hopkins University, Homewood
      Campus, Baltimore, MD 21218\\
      e-mail: heckman@pha.jhu.edu}

\begin{abstract}

We have extended the evolutionary synthesis models by Leitherer et al. (1999b)
by including a new library of B stars generated from the IUE high-dispersion 
spectra archive. We present the library and show how the stellar 
spectral properties vary according to luminosity classes and spectral types. 
We have generated synthetic UV spectra for prototypical young stellar populations varying the IMF and the star 
formation law. Clear signs of age effects are seen in all models. The contribution 
of B stars in the UV  line spectrum is clearly detected, in particular for greater ages
 when O stars have evolved. With the addition of the new library we are able to 
 investigate the fraction of stellar and interstellar contributions and the variation in the 
spectral shapes of intense lines. We have used our models to date the spectrum 
of the local super star cluster NGC~1705--1. Photospheric lines of C~III~$\lambda$1247, Si~III~$\lambda$1417, 
and  S~V~$\lambda$1502 were used as diagnostics to date the burst of NGC 1705--1 at 10 Myr. 
Interstellar lines are clearly seen in the NGC 1705--1 spectrum. Broadening and blueshifts 
of several resonance lines are stronger in the galaxy spectrum than in our models and are 
confirmed to be intrinsic of the galaxy. Si~II~$\lambda$1261 and Al~II
$\lambda$1671 were found to be pure interstellar 
lines with an average blueshift of 78~km~s$^{-1}$ due to a directed outflow of 
the interstellar medium. We have selected the star-forming galaxy 1512--cB58 as 
a first application of the new models to high-$\it{z}$ galaxies. This galaxy is 
at z~=~2.723, it is gravitationally lensed, and its high signal-to-noise Keck
spectrum show features typical of local starburst galaxies, such as  NGC 1705--1.
Models with continuous star formation were found to be more adequate for 
1512--cB58 since there are spectral features typical of a composite stellar population 
of O and B stars. A model with $Z =$~0.4~Z$_{\odot}$ and an IMF with $\alpha$~=~2.8 
reproduces the stellar features of the 1512--cB58 spectrum. 
\end{abstract}



\bigskip
\bigskip

\centerline{\it ApJ in press - February 10, 2000 - vol. 530}

\section{Introduction}

O and B stars with zero-age-main-sequence (ZAMS) masses~$\ge$~5 M$_{\odot}$ are 
the natural tracers 
of recent or current star-forming activity. These stars play a fundamental role
in the evolution of a galaxy since they are
responsible for the release of ionizing radiation, chemically enriched material, 
and kinetic energy 
via winds and supernovae (Maeder \& Conti 1994). 
The most prominent stellar emission and absorption lines in the spectra of O and 
B
stars are in the ultraviolet (UV) region. An old underlying stellar population 
and strong nebular
emission lines may mask the contribution of a hot stellar population in the 
optical. 
Therefore, the preferred wavelength
region to detect massive stars is longward of the Lyman $\alpha$ forest but 
short enough to
minimize spectral contamination from older stellar populations; i.e.
1216~{\AA}~$\le$~$\lambda$~$\le$~2500~{\AA}. Although this spectral region 
cannot be 
observed from the ground, the amount of high quality data obtained with the {\it 
International
Ultraviolet Explorer } (IUE) has led to detailed spectral analyses of massive 
stars in the UV. 
The IUE spectral atlases of massive stars of Walborn, Nichols-Bohlin, \& Panek
(1985), Rountree \& Sonneborn (1993) and Walborn, Parker, \& Nichols (1995b) 
provide a unique tool for the
spectral classification of O and B stars. Moreover, the IUE spectra have been 
extensively used 
to test stellar atmosphere models (e.g., Groenewegen \& Lamers 1989).

Starbursts are a major mode of high-mass star formation. In the local universe 
four 
bright starburst galaxies (M 82, NGC 253, NGC 4945 and M 83) account for about 
25\% of the 
high-mass star formation rate (Heckman 1997). IUE, together with the Hubble 
Space Telescope (HST), and the Hopkins 
Ultraviolet Telescope (HUT) satellites, has
dramatically increased our understanding of the 
starburst phenomenon (e.g., Huchra et al. 1983; Fanelli, O'Connell, \& Thuan et 
al. 1988; Kinney et al.
1993; Leitherer et al. 1995a, 1996). 
The UV region of the starburst spectra is dominated by strong resonance 
transitions of, e.g., 
Si~IV~$\lambda$1400 and C~IV~$\lambda$1550, which are the strongest spectral 
features of 
massive stars in the satellite--UV. The main contributors to the UV flux are 
 O and early B stars. O stars dominate during the earliest starburst phase, when the 
spectra
are characterized by broad ($\sim$ 2000 km s$^{-1}$), highly ionized ($\sim$ 50 
eV) stellar wind lines.
However, after a few million years O stars disappear, and B stars become the 
main source of the UV flux 
in the integrated spectrum. Therefore, in order to analyze the stellar content 
of star-forming 
galaxies it is necessary to understand the contribution not only of O but also 
of B stars to the 
integrated spectra of these objects. 

Evolutionary synthesis models were used by us before to analyze the integrated 
light of galaxies and to 
understand the hot--star content of local starburst galaxies (Robert, Leitherer, 
\& Heckman 1993; Leitherer et al. 1996; 
Heckman \& Leitherer 1997). We used Galactic O- and Wolf-Rayet stars as building 
blocks to synthesize 
the ultraviolet spectra of starburst galaxies in the local universe. A stellar 
library from IUE high-dispersion 
spectra was created and implemented into a synthesis code to model the 
ultraviolet spectra of O-star dominated
starburst galaxies. The limitation 
of the synthesis technique is due to the uncertainties of the stellar evolution 
models and the quality of 
the stellar libraries used. Recently, in Leitherer et al. (1999b, hereafter 
Starburst99) 
we presented an upgrade of the models of Leitherer \& Heckman (1995b). 
Starburst99 includes the newest model atmosphere grid of Lejeune, Cuisinier, \& 
Buser (1997) 
and the latest Geneva evolutionary tracks. 
In this paper we present a new addition to Starburst99: a library of IUE 
high-dispersion 
spectra of B stars. With the addition of this new library 
we are able to synthesize high-resolution (0.75~{\AA}) UV spectra that are 
typical of more evolved starbursts ({\it t}~$>$~10~Myr) 
dominated by B stars. We can also detect the contribution of B stars in young 
mixed populations which
have high--ionization lines formed in O star winds but also have low--ionization 
lines from B supergiants.
Section II describes the computational method. Section III introduces the new 
library of B stars and 
identifies the main diagnostic lines. Models for stellar population with different star formation laws 
and initial mass functions (IMF) are discussed in 
Section IV. Section V presents a set of synthetic models generated to reproduce 
the UV spectrum of the super star cluster in
the local starburst NGC~1705, and Section VI discusses the stellar population of 
the high-$\it{z}$ galaxy 1512--cB58. 
Conclusions are presented in Section VII.

\section{Method}

Stars are generated in Starburst99 according to a star-formation 
law as a function of time and distributed
along the ZAMS with a power-law IMF. The
stars are evolved from the ZAMS using the evolutionary models of the Geneva 
group.  
Two types of star-formation law are considered: (i) an instantaneous burst with 
no 
subsequent star formation
and (ii) continuous star-formation at a constant rate. The stellar IMF is 
approximated by a power-law with an
exponent $\alpha$ and a lower and upper mass-limit $M_{\rm{low}}$ and 
$M_{\rm{up}}$, respectively. 
The choices of IMFs are the Salpeter IMF with $\alpha$=2.35, and an approximate 
Miller-Scalo IMF with $\alpha$~=~3.3
(Salpeter 1955; Miller \& Scalo 1979). $M_{\rm{low}}$ is
assumed to be 1~M$_{\odot}$ and $M_{\rm{up}}$~=~100 M$_{\odot}$. A truncated 
Salpeter IMF with 
$M_{\rm{low}}$~=~1~M$_{\odot}$ and $M_{\rm{up}}$~=~30 M$_{\odot}$ is also 
available in Starburst99 but 
was not used in this work. We refer to Leitherer \& Heckman
(1995b) and Leitherer et al. (1999b) for more details on each individual 
parameter.
Chemical evolution is not treated self-consistently in our models and each 
stellar
generation has the same metallicity during the evolution of the population. 

The proper transformation from the theoretical (effective temperature 
$T_{\rm{eff}}$; luminosity $L$) to the 
observed (spectral type) plane is a concern in any evolutionary synthesis model. 
Recall that the theoretical
prediction is $T_{\rm{eff}}$ and $L$,  whereas the observed spectra are assigned 
via their corresponding spectral types. 
Therefore, the adopted $T_{\rm{eff}}$ vs. spectral type calibration becomes 
crucial. Starburst99 assigns spectral types to
positions in the Hertzsprung-Russell diagram (HRD) using Schmidt-Kaler's (1982) 
calibration. The choice of this
calibration was validated by the success in reproducing both the integrated UV 
spectrum as well as the spectral type
distribution of the `starburst Rosetta Stone' 30 Doradus (Vacca et al. 1995). 
Schmidt-Kaler's
calibration has also been shown to agree with more recent calibration methods 
such as those of Chlebowski \& Garmany (1991) for O-type stars and 
Humphreys \& McElroy (1984) for later types (Massey et al. 1995).

B-supergiant dominated phases add a new challenge to the modeling: evolution 
proceeds very rapidly in those
parts of the HRD where luminous B stars are located, and the star counts may be 
inaccurate if the interpolation scheme
fails. Our scheme performs a nearest neighbor search in the $L$ vs. 
$T_{\rm{eff}}$ plane to assign spectral types. We can test
the reliability of our method by requiring that a constant star-formation model 
correctly reproduces the observed
distribution of spectral types in the solar neighborhood within a few kpc from 
the Sun. If a sufficiently large volume is 
sampled, like for $d$~$\ge$~1 kpc, local variations
of the star-formation rate are expected to smooth out, and differences between 
theoretical and observed stellar statistics
are primarily due to errors in the evolutionary model lifetimes, the spectral 
calibration, and the interpolation scheme.
We performed this test by comparing the stellar population generated by our code 
with the observed number of massive stars compiled by Blaha \& Humphreys (1989, 
BH hereafter). 
Their compilation includes luminous O and B stars and supergiants of all 
spectral types. 
Figure~1 of BH shows the HRD derived for stars in 91 associations and 
clusters within 3 kpc of the Sun. We used this figure to determine the number of 
stars for each 
spectral type. However, as shown by the luminosity function 
obtained by BH, the data are severely affected by incompleteness at 
$M_{V}$$>$--6 
(see Figure~8 in BH). Therefore, we
have to take incompleteness into account before we compare the numbers predicted 
by our models with the numbers
given in BH. We applied an incompleteness correction to the BH data in the 
following way. A first-order polynomial 
with slope = +0.71 was fitted to the luminosity function at 
--9~$\le$~$M_{V}$~$\le$~--7 where the data are still complete. 
The fit gives the expected number of stars for each $M_{V}$. The incompleteness 
factor for magnitude
bins $M_{V}$$>$--7
was determined by an extrapolation of the fitted line up to $M_{V}$$\sim$--4 ($M_{V}$~=~--4.5 
for O9V stars). 
Applying this correction factor to BH's data (using again  
Schmidt-Kaler's calibration for the conversion from $M_{V}$ to spectral type) we 
calculated the corrected number of 
stars for each spectral type. Table 1 shows the number of stars that were 
classified as OV and BI 
from BH's data before and after correcting for incompleteness. Note the severe 
incompleteness of O main-sequence stars of
all temperatures. In contrast, the visually brighter B supergiants of BH's 
sample are almost complete.
Table 1 also shows the predicted number of OV and BI 
stars generated by a model with Salpeter IMF, continuous star-formation law, and 
age = 20~Myr. 
The predicted spectral types become age-independent at this age since an 
equilibrium between star formation and stellar death
has been reached. The observed and computed ratios OV/BI are 6.9 and 5.3, 
respectively. We did not attempt to optimize the
agreement further, e.g. by varying the IMF, since the uncertainties due to the 
incompleteness correction by far exceed the 
35\% difference between the model and the observations. 
While the overall observed ratio of evolved over unevolved massive stars is
reproduced rather well by our models, the detailed supergiant distribution is
not. The observed HRD shows a nearly flat distribution from early- to late-B 
supergiants, whereas the models predict an initial decrease and a subsequent
increase for late B supergiants. This reflects the uncertainties of the
details of the evolution models in these particular HRD regions. One should
be aware that synthetic spectra showing mid- and late-B supergiant features
are affected by this uncertainty. In practice, this should rarely be a concern.
For an evolved single starburst population, the spectrum will be dominated by stars
close to the main-sequence turn-off, and not by mid- to late-B supergiants.
For a mixed population, mid- to late-B supergiants are present, but they
hardly contribute to the spectrum due to their low UV flux. Nevertheless this
caveat should be kept in mind. 
We conclude that our 
adopted $T_{\rm{eff}}$ vs. spectral type
interface for B supergiants works reasonably well.
 
\begin{table*}
\caption{Number of Stars for Each Spectral Type \label{table:observed}}
\begin{tabular}{cccccccc} \hline\hline
\multicolumn{1}{c}{Spectral} & \multicolumn{3}{c}{Number of Stars} 
&\multicolumn{1}{c}{Spectral}& \multicolumn{3}{c}{Number of Stars}
\\ \cline{2-4} \cline{6-8}
\multicolumn{1}{c}{Type} & \multicolumn{1}{c}{Observed} 
&\multicolumn{1}{c}{Corrected} &\multicolumn{1}{c}{Predicted}
& \multicolumn{1}{c}{Type} & \multicolumn{1}{c}{Observed}
&\multicolumn{1}{c}{Corrected} &\multicolumn{1}{c}{Predicted}\\
\hline
O4V   & 2  & 3   & 201 & B0I   & 27 & 43 & 633 \\
O5V   & 8  & 26  & 411 & B1I   & 32 & 51 & 668 \\
O6V   & 30 & 120 & 1138 & B2I   & 54 & 86& 19 \\
O7V   & 63 & 416 & 1730 & B3I   & 28 & 45& 15 \\
O8V   & 99 & 792 & 3300 & B5I   & 23 & 35& 8 \\
O9V   & 43 & 645 & 8880 & B7I   & 2  & 2& 4 \\
     &     &     &      & B8I   & 6  & 3& 4 \\
     &     &     &      & B9I   & 8  & 13& 518 \\
\hline
Total& 245 & 2002& 15660&       & 180 & 278& 2974 \\
\hline
\end{tabular}
\end{table*}
 

\section{UV lines in O and B stars}

O and B stars span a wide range of physical properties such as $T_{\rm{eff}}$, 
$L$, radius, and mass. O stars have $T_{\rm{eff}}$ 
in the range of 3--5$\times$10$^{4}$ K whereas B stars vary from 
1--3$\times$10$^{4}$ K. Both are
highly luminous with luminosities for O and B stars varying between 
2$\times$10$^{6}$--2$\times$10$^{5}$~L$_{\odot}$ 
and 1$\times$10$^{5}$--2$\times$10$^{2}$~L$_{\odot}$, respectively. Their 
luminosities are high enough to
drive massive outflows in the form of radiation-driven stellar winds. 
The stellar radiation field below 912~{\AA} is the driving force of a wind 
(Lucy \& Solomon 1970; Castor, Abbott, \& Klein 1975). As the fraction of the stellar ionizing luminosity over 
the bolometric luminosity decreases from hot 
O stars to cooler B stars, the wind momentum decreases as well. Typical wind 
velocities for O stars are 
2--3 $\times$10$^{3}$ km s$^{-1}$, whereas for B stars the values are up to 10 
times lower. At the
same time, mass-loss rates decrease by at least an order of magnitude from 
spectral types O to B
(e.g. Bieging, Abbott, \& Churchwell 1989). Different wind properties and a 
softer ionizing radiation
field dramatically modify the ultraviolet spectral morphology of B with respect 
to O stars: typical
strong ultraviolet wind lines have lower ionization potentials in the 20 eV 
range (vs. 50 eV 
for O stars) and line widths and shifts of a few {\AA} (vs 10~{\AA} for O 
stars). Note that 
despite lower wind densities, B-star wind lines are still optically thick and 
therefore are 
strong features. 

Typical UV spectra of O and B stars show three types of lines: stellar 
photospheric 
lines, stellar-wind lines, and interstellar absorption lines. The most common 
ionized photospheric absorption lines are from C, N, O, Si, and Fe of low and high 
excitation potentials. Pure {\em photospheric lines} are formed in layers which 
are almost in
hydrostatic equilibrium ($v~\ltsim~10^{2}~$km~s$^{-1}$) and are not affected by stellar wind outflows.
Since these lines originate from excited levels, they do not form in stellar 
winds or in the interstellar medium (ISM).
Therefore, they are a unique tracer of stars eliminating the uncertainty of 
contamination by an interstellar contribution.
However, high-quality spectra are required for detection, as they are generally 
weak. The most notable exception
is N~IV~$\lambda$1720 which forms in O-star winds despite originating from a 
highly excited level. It
originates from the level $^{1}$P$_{0}$ which is connected to the ground state 
$^{1}$S by an allowed line at
$\lambda$765~{\AA}. This line is optically thick, so that the $^{1}$P$_{0}$ 
state behaves as ground state.
For very hot stars for which the Lyman continuum is less optically thick, the 
state will be 
depopulated by the photon escape from the resonance line at $\lambda$765~{\AA}. 
Strong resonance lines of any atomic and
ionic species can form in {\it stellar winds} and in the {\it ISM}. Generally, 
species from lower ionization potentials are stronger in the
ISM and lines from higher ionization potentials are stronger in winds. While 
this distinction is almost always true in O
stars, it does not apply to B stars where it is often difficult to distinguish 
the stellar and interstellar
components. Defining the contributions from the photosphere, wind, and ISM in B stars is 
one of the major goals of this paper. 

Tables 2 and 3 summarize the main stellar and interstellar lines found in the 
spectra of O and B stars. 
The atomic data are from Moore (1950), except for C~III~$\lambda$1426.45, 1427.84 
and S~V~$\lambda$1501.76 which
were taken from van Hoof's line list at {\tt 
http://www.pa.uky. edu/\~\,peter/atomic}. The general trend in 
these tables is a decrease of the ionization potential of the
dominant line with decreasing $T_{\rm{eff}}$ and with increasing contribution 
from the ISM.
Tables 2 and 3 are intended to give the general trends in the spectral 
morphology of OB stars. Very
often it is difficult to disentangle the stellar and interstellar components of 
a resonance line. An
example is Si~IV~$\lambda$1393.73, 1402.73 (also called Si~IV~$\lambda$1400). This 
doublet has a strong wind 
contribution in giants and supergiants,
and an additional weaker interstellar line. The stellar and interstellar 
components are usually easy
to distinguish due to their large velocity separation. Si~IV~$\lambda$1400 is 
also a strong (unshifted)
absorption line in OB main-sequence stars, reaching maximum strength around B1V. 
The lines become
progressively weaker towards earlier and later types, with a corresponding 
increase in line strength
of Si~V and Si~III transitions, respectively. The weak Si~IV~$\lambda$1400 lines in 
the earliest O
main-sequence stars are almost always interstellar, yet at IUE resolution their profiles are 
indistinguishable from those
of photospheric lines. This degeneracy may be less of an issue when spectral 
morphology is concerned.
It does, however, become crucial when Galactic stars are used as templates for 
comparison with
starburst galaxies. The relative strengths of stellar and interstellar lines may 
well be different in
those galaxies due to different ISM conditions. For this reason we attempted to 
describe the morphology
of the stellar and interstellar lines separately in Tables 2 and 3.

\tiny
\begin{table*}
\caption{Main OB Stellar Features}
\begin{tabular}{llrr|l} \hline\hline
\multicolumn{1}{c}{Ion}  & \multicolumn{1}{c}{$\lambda$} & 
\multicolumn{1}{c}{E.P.} &
\multicolumn{1}{c}{I.P.} &
\multicolumn{1}{c}{Comments}\\
&\multicolumn{1}{c}{({\AA})}& \multicolumn{1}{c}{(eV)$_{\rm{low}}$} &
\multicolumn{1}{c}{(eV)$_{\rm{low}}$} & \\
\hline
N~V   & 1238.80   & 0.00 & 77.47 & O3V--O6V, O3III--O9IIII, O3I--O9I \\  
      &           &      &       & B0III, B0I, B1I - wind \\
      & 1242.78   & 0.00 & 77.47 & see N~V~$\lambda$1238.80 \\
C~III  & 1247.37   & 12.64& 24.38 & O3V--O9V, O3III--O9III, O3I--O9I; stronger in 
$\ge$~O7\\
      &           &      &       & B0V--B8V, B0III--B8III, B0I--B8I; stronger in 
$\le$~B2\\
S~II   & 1250.50   & 0.00 & 10.36 & IS in O stars; B0I--B8I stronger in 
$\ge$~B5\\
      & 1253.79   & 0.00 & 10.36 & see S~II $\lambda$1250.50\\
      & 1259.53   & 0.00 & 10.36 & close to Si~II $\lambda$1260.66\\
Si~II  & 1260.66   & 0.00 & 8.15  & purely IS in O stars\\
      &           &      &       & B0V--B8V, B0III--B8III, B0I--B8I - blended 
with IS\\
      & 1265.04   & 0.04 & 8.15  & B1V--B8V, B1III--B8III, B1I--B8I; stronger in 
$\ge$~B3\\
Si~III & 1294.55   & 6.51 & 16.34 & O stars show Fe~II in this region \\
      &           &      &       & B0V--B8V, B0III--B5III, B0I--B8I \\
      & 1296.72   & 6.50 & 16.34 & B0V--B8V, B0III--B3III, B0I--B8I - stronger 
in BIs \\
      & 1298.90   & 6.55 & 16.34 & B0V--B8V, B0III--B8III, B0I--B8I - stronger 
in $\ge$~B2\\
Si~II  & 1304.41   & 0.00 & 8.15  & B3V--B8V, B3III--B8III, B3I--B8I - blended 
with IS, wind\\
      & 1309.28   & 0.04 & 8.15  & B3V--B8V, B3III--B8III, B5I--B8I - blended 
with IS, wind\\
C~II   & 1334.52   & 0.00 & 11.26 & blended with C~II $\lambda$1335.68 - IS\\
      & 1335.68   & 0.01 & 11.26 & IS in O stars; B0V--B8V, blended with IS in 
$\ge$~B3V\\
      &           &      &       & B0III--B8III, blended with IS in 
$\ge$~B5III\\
      &           &      &       & B0I two components (IS C~II 
$\lambda$1334.52)\\
      &           &      &       & B1I--B8I blended with IS and show wind\\
Si~IV  & 1393.73   & 0.00 & 33.49 & O3V--O9V, O3III--O5III, O3I - no wind\\
      &           &      &       & O6III--O9III wind, O4I--O9I - strong wind\\
      &           &      &       & B0V--B8V stronger $\le$ B3V,B0III--B3III wind 
in B0III\\
      &           &      &       & B0I--B8I stronger $\le$ B3I - strong wind\\ 
      & 1402.73   & 0.00 & 33.49 & see Si~IV $\lambda$1393.73\\
Si~III & 1417.20   & 10.23& 16.34 & O stars show Fe~II lines in this region\\
      &           &      &       & B2V--B8V, B6III--B8III, B0I--B8I stronger in 
$\le$~B2I\\
\hline
\end{tabular}
\end{table*}
\normalsize

\tiny
\begin{table*}
\addtocounter{table}{-1}
\caption{Main OB Stellar Features - Cont.}
\begin{tabular}{llrr|l} \hline\hline
\multicolumn{1}{c}{Ion}  & \multicolumn{1}{c}{$\lambda$} & 
\multicolumn{1}{c}{E.P.} &
\multicolumn{1}{c}{I.P.} &
\multicolumn{1}{c}{Comments}\\
&\multicolumn{1}{c}{({\AA})}& \multicolumn{1}{c}{(eV)$_{\rm{low}}$} &
\multicolumn{1}{c}{(eV)$_{\rm{low}}$} & \\
\hline
C~III  & 1426.45   & 29.53& 24.38 & close to Fe~V $\lambda$1429,1430 in O stars\\
      &           &      &       & B0V, B1V, B0I--B2I, B0III--B2III\\
      & 1427.84   & 29.53& 24.38 & see C~III $\lambda$ 1426.45\\
Si~II  & 1485.40   & 6.83 & 8.15  & close to N~IV] in O stars\\
      &           &      &       & B3V--B8V, B3III--B8III, B5I--B8I - weaker in 
BIs\\
S~V    & 1501.76   & 15.76& 47.30 & O3V--O9V, O3III narrow, O5III--O9III, 
O3I--O9I\\
      &           &      &       & B0V, B1V, B0I, B1I\\
Si~II  & 1526.70   & 0.00 & 8.15  & O5III--O9III, two blended components\\
      &           &      &       & B0V--B8V, B0III--B8III, B0I--B8I, two blended 
components\\
      & 1533.44   & 0.04 & 8.15  & O stars show Fe~IV in this region\\
      &           &      &       & B0V--B3V two components, B4V--B8V only IS\\
      &           &      &       & B0III--B3III two components, B5III--B8III 
only IS\\
      &           &      &       & B0I--B8I two blended components\\
C~IV   & 1548.19   & 0.00 & 47.89 & O3V--O7V, O3I--O5I - strong wind\\
      &           &      &       & O3III--O9III, O6I--O9I\\
      & 1550.77   & 0.00 & 47.89 & B0V, B0III, B0I--B3I - strong wind\\
Fe~II  & 1608.45   & 0.00 &  7.87 & O3V, O4V, O3III--O7III - two blended 
components\\
      &           &      &       & $\ge$~O5V, O3I--O9I - only IS\\
      &           &      &       & B0V--B8V, B0III--B8III, BI--B8I, two blended 
components\\
He~II  & 1640.49   & 40.64& 24.59 & O3V--O9V, O3III--O9III; stronger in 
$\ge$~O7\\
      &           &      &       & O3I--O5I - wind, O6I--O9I\\
      &           &      &       & B0V, B1V, B0III, B1III, B0I--B3I\\
Al~II  &  1670.81  & 0.00 & 5.99 & IS in O stars\\
      &           &      &       & B0V--B8V stronger $\ge$B5V, 
B5III--B8III,B1I--B8I\\
N~IV   &  1718.52  & 16.13& 47.45 & similar to Si~IV $\lambda$1400 in O stars\\
      &           &      &       & close to Al~II/III blend in B stars\\
\hline
\end{tabular}
\end{table*}

\begin{table*}
\caption{Main Interstellar Features}
\begin{tabular}{llll} \hline\hline
\multicolumn{1}{c}{Ion}  & \multicolumn{1}{c}{$\lambda$} & 
\multicolumn{1}{c}{E.P.} &
\multicolumn{1}{c}{I.P.} \\
&\multicolumn{1}{c}{({\AA})}& \multicolumn{1}{c}{(eV)$_{\rm{low}}$} &
\multicolumn{1}{c}{(eV)$_{\rm{low}}$}  \\
\hline
N~V   & 1238.80 & 0.00 & 77.47 \\
      & 1242.78 & 0.00 & 77.47\\
S~II   & 1250.50 & 0.00 & 10.36\\
      & 1253.79 & 0.00 & 10.36\\
      & 1259.53 & 0.00 & 10.36\\
Si~II  & 1260.66 & 0.00 & 8.15\\
O~I   & 1302.17 & 0.00 & 0.00\\
Si~II  & 1304.41   & 0.00 & 8.15\\
C~II   & 1334.52   & 0.00 & 11.26\\
      & 1335.68   & 0.01 & 11.26\\
Si~IV  & 1393.73   & 0.00 & 33.49\\
      & 1402.73   & 0.00 & 33.49\\
Si~II  & 1526.70   & 0.00 & 8.15\\
C~IV   & 1548.19   & 0.00 & 47.89\\
      & 1550.77   & 0.00 &47.89\\
Fe~II  & 1608.45   & 0.00 & 7.87\\
Al~II  & 1670.81   & 0.00 & 5.99\\
\hline
\end{tabular}
\end{table*}

\normalsize

Starburst99 includes a high-dispersion library of stellar
types O and Wolf-Rayet and a low-dispersion library of B stars (see Robert et 
al. 1993 for a 
description of the libraries). This means that synthetic spectra generated to 
reproduce evolved 
bursts are composed mainly of low resolution ($\sim$ 6~{\AA}) B star spectra 
which do not give enough
detail for the line profiles. Therefore, in order to date an evolved burst and 
to study the stellar content of 
star-forming galaxies, a library of high-dispersion IUE spectra of B stars is 
needed.
The inclusion of this library will also allow the study of the IMF in the 
intermediate-mass range. For example, 
a B8V star corresponds to a main-sequence 
mass of $\sim$ 3 M$_{\odot}$ with a lifetime of about 0.5 Gyr (Schaller et al. 
1992) whereas
O stars have ZAMS masses above $\sim$ 20 M$_{\odot}$ and lifetimes of less than 
20~Myr. 

We have used the atlases of IUE high-dispersion spectra by Rountree \& Sonneborn 
(1993) and
Walborn et al. (1995b), and the sample by Howarth et al. (1997) as guidelines to
select a library of B stars. Although these atlases are available in electronic 
version, these authors have only
published rectified spectra. In order to be consistent with our previously 
generated library we
have performed an independent extraction from the IUE archive and  applied the 
same rules to the B-star library as 
we did in the case of O and WR stars. We have extracted high-dispersion, SWP, 
spectra from the IUE archive for all stars that showed no peculiarities in their 
spectra and
that had IUE data from those atlases. The standard IUE processed data were 
considered to 
give reasonable results and no further calibration processing was needed. IDL was 
used to combine the 
echelle orders and the task $\it {scombine}$ in IRAF to sum all the spectra of 
the same star. 
The number of spectra used in each star depended on the number of spectra 
available in the archive
and their quality. For example, 10 spectra of HD 58350 (exposure times from 40~s 
to 120~s) were extracted from the IUE archive.
After inspecting each spectrum, they were considered to show no peculiarity and 
all of them were used in
the final summing. HD 79186, however, had only three spectra in the archive 
(exposure times 1200~s, 2520~s, and 3300~s) but
they were considered of quality similar to HD 58350 and were used in the final 
summing.

Each spectrum was continuum normalized by dividing the calibrated spectra by a 
spline fit to the 
continuum. The task $\it {continuum}$ in IRAF was used to perform the fit. This
continuum normalization technique reveals prominent features such as a 
depression 
around 1400~{\AA} and 1600~{\AA}. As pointed out in Robert et al. (1999), these 
depressions 
are actually real features and are caused by Fe~II , Fe~III, Fe~IV and Fe~V lines. 
These Fe ions are diagnostics
of massive stars. Very few of them are transitions from the ground state which 
make them unlikely to be formed in the interstellar medium. 

\begin{figure}[h]
\plotone{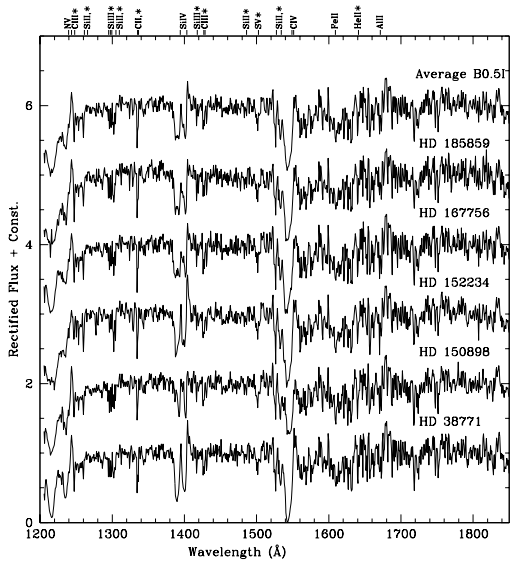}
\caption{Rectified spectra of B0.5I stars and the average spectrum. 
Star names are given on the right corner of each spectrum. Excited lines 
(photospheric) are marked with ``$\ast$''. ``,$\ast$'' is used when only the
second component is an excited line. See Table 4 for a list of stars. The 
spectra are rectified.} \label{f1}
\end{figure}

The final library was built with 
73 B stars of spectral types B0--B8 and luminosity classes I, III, V (Table 4). 
We generated average spectra for each spectral type and luminosity class. 
For example, the class B0.5I spectrum 
is made by averaging the spectra of HD 38771, HD 150898, HD 152234, HD 167756,
and HD 185859. Figure~\ref{f1} illustrates the differences between individual 
stars and the average spectrum. Overall the average spectrum preserves the main
features common in each individual spectrum. Features which are 
not strong in all stars are smoothed out in the average spectrum. Therefore, one should
take into account that using average spectra may create uncertainties in
the final generated spectrum. However, the average spectra suit the qualitative application of our method. Classes II and IV were not taken from the IUE archive. 
They were created by taking the average between the nearest luminosity classes; 
i.e. IV=(III+V)/2 and II=(I+III)/2.

\begin{table*}
\caption{B Stars - IUE Library}
\begin{tabular}{llllll} \hline\hline
\multicolumn{1}{c}{Spectral}&\multicolumn{1}{c}{Stars}
&\multicolumn{1}{c}{Spectral} & \multicolumn{1}{c}{Stars}
&\multicolumn{1}{c}{Spectral}&\multicolumn{1}{c}{Stars} \\ 
\multicolumn{1}{c}{Type} & & \multicolumn{1}{c}{Type} & 
&\multicolumn{1}{c}{Type} \\
\hline
B0V       &HD 53755  &  B0III       &HD 48434& B0I       &HD 37128 \\
          &HD 143275 &              &        &           &HD 91968 \\
          &HD 149438 &              &        &           &HD 122879 \\		
B0.5V     &HD 55857  &  B0.5III  &HD 218376& B0.5I       &HD 38771 \\
	  &HD 144217 &		 &HD 332407&		 &HD 150898 \\
 	  & 	     &		 &         &		 &HD 152234 \\
 	  & 	     &		 &         &		 &HD 167756 \\
 	  & 	     &		 &         &		 &HD 185859  \\
B1V       &HD 31726  & B1III     &HD 23180 & B1I         &HD 106343 \\ 
          &HD 37018  &		 &HD 147165&		 &HD 148422 \\	 
          &HD 144470 &		 &         &		 &HD 148688 \\
 	  & 	     &		 &         &		 &HD 150168 \\
 	  & 	     &		 &         &		 &HD 163522 \\
 	  & 	     &		 &         &		 &HD 190603 \\
B1.5V     &HD 35299  &           &         &             & \\
          &HD 36959  &		 &	   &		 & \\
B2V       &HD 42401  &  B2III    &HD 35468 & B2I          &HD 41117 \\ 
          &HD 64802  &	         &	   & 	          &HD 42087 \\
 	  & 	     &	         &	   & 	          &HD 92964 \\
 	  & 	     &	         &	   & 	          &HD 116084 \\
B2.5V     &HD 148605 &           &         &             & \\
          &HD 175191 &		 &	   &		 & \\
B3V       &HD 32630  & B3III         &HD 79447& B3I     &HD 53138 \\
          &HD 74280  &  	    &	    &	       &HD 75149 \\
          &HD 120315 &  	    &	    &	       &HD 111973 \\
          &HD 190993 &  	    &	    &	       &HD 225093 \\
B4V       &HD 20809   &  	    &       &          &          \\
          &HD 65904   &  	    &       &          &          \\
B5V       &HD 25340   & B5III       &HD 22928& B5I       &HD 58350 \\
          &HD 34759   &  	    &HD 34503&	         &HD 79186 \\
          &HD 188665  & 	    &        &	         &HD 1000943 \\
B6V       &HD 90994   & B6III    &HD 23302& 	         &	    \\
B7V       &HD 17081   &  	 & 	 &		 & \\
          &HD 29335   &  	 & 	 &		 & \\
          &HD 87901   &  	 & 	 &		 & \\	  
B8V       &HD 23324   & B8III    &HD 23850 & B8I       &HD 34085  \\
\hline
\end{tabular}
\end{table*}

\begin{figure}[h]
\plotone{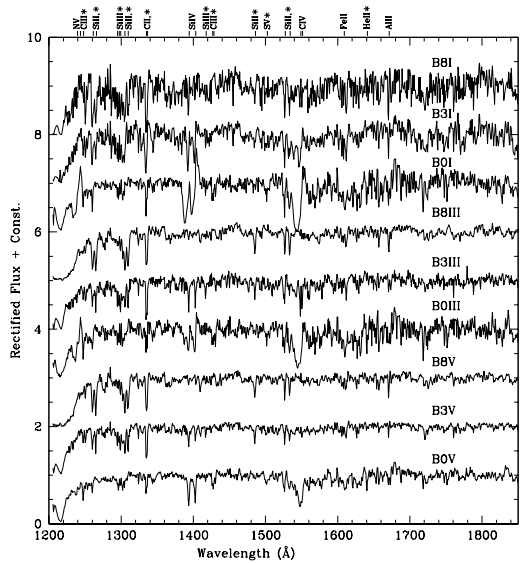}
\caption{Rectified spectra of representative B stars. Spectral types are given on 
the 
right corner of each spectrum. Spectral types with more than one spectrum were
averaged. Excited lines (photospheric) are marked with ``$\ast$''. ``,$\ast$'' is used when only the
second component is an excited line. See Table 4 for a list of stars. The
spectra are rectified.} \label{f2}
\end{figure}

Figure~\ref{f2} shows representative spectra of our library. It is important 
to notice that most of the
features seen in the IUE stellar spectra are real physical lines and not noise 
(Nemry, Surdej, \&
Hernaiz 1991). Several features are important diagnostics in B stars. Silicon, 
for example, 
has been shown to be a good diagnostic for the evolutionary state of a star 
because the 
atmospheric abundance is expected to remain unchanged by nuclear processing 
(e.g. Walborn 1971;
Prinja 1990). In the following we will discuss those diagnostics in more detail 
which have
the greatest potential in constraining extragalactic populations.

\subsection{Silicon lines}

\begin{figure}[h]
\plotone{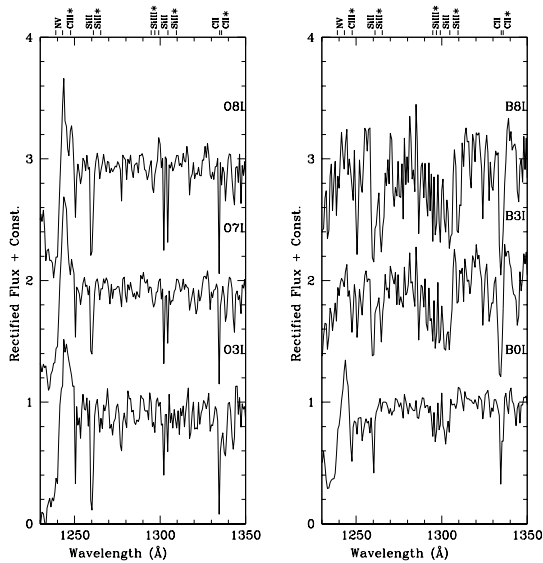}
\caption{Spectral region around 1300~{\AA} in O (left) and B 
(right) supergiants. The strongest features are labeled. Excited lines 
(photospheric) are marked with ``$\ast$''.} \label{f3}
\end{figure}

Si~II~$\lambda$1260.66, 1265.04 are found in B stars of all
luminosity classes. Si~II~$\lambda$1260.66 is a strong interstellar line in O and 
B stars 
(Figure~\ref{f3}). Si~II~$\lambda$1265.04 is photospheric and is not seen 
in O stars. The
irregular shape of Si~II~$\lambda$1265.04 suggests the presence of Si~II~$\lambda$1264.73 
but with the spectral resolution (0.75~{\AA}) of our library 
we cannot separate the two lines.

The Si~III~$\lambda$1295--1300 photospheric lines are formed by six lines in the multiplet 
$^{3}$P$^{0}$-$^{3}$P. Three lines, Si~III~$\lambda$1294.55, $\lambda$1296.72, and 
$\lambda$1298.90 are found in B stars of 
all luminosity
classes. Si~III increases in strength through late B supergiants and it
is absent in O supergiants (Figure~\ref{f3}).  The blending with  
O~I$\lambda$1302.17 (pure interstellar line), Si~II~$\lambda$1304.41, and Si~II 
$\lambda$1309.28 
enhances the Si~III strength in late type stars if the 1300~{\AA} region is 
observed at low spectral resolution.

\begin{figure}[h]
\plotone{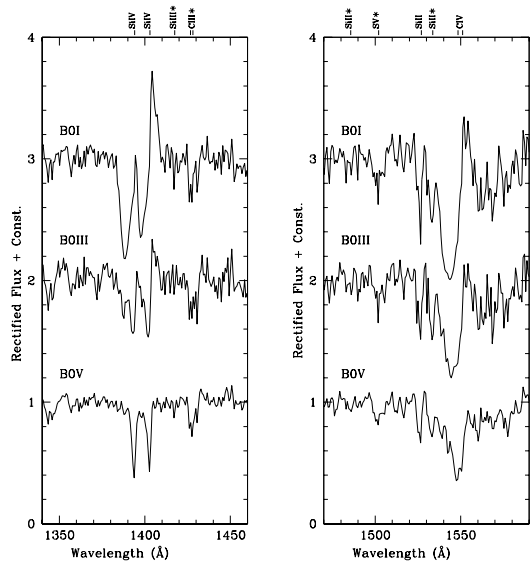}
\caption{Spectral region around 1400~{\AA} (left) and 
1525~{\AA} (right) in B0 stars. Excited lines (photospheric) are marked with
``$\ast$''.} \label{f4}
\end{figure}

The Si~IV~$\lambda$1393.73, 1402.73 resonance line doublet develops from 
photospheric plus 
interstellar in the luminosity class V spectrum, through intermediate wind 
profiles at class III,
to a full P~Cygni profile in the luminosity class I. This is due to the fact 
that there is a correlation 
between luminosity and wind density. Only luminous stars have stellar winds 
dense enough to generate 
enough opacity to produce a strong P~Cygni profile. Figure~\ref{f4} 
shows the behavior of 
these lines for early B stars. The change in the Si~IV~$\lambda$1400 profile with  
luminosity class makes it a powerful diagnostic of B stars. These lines are 
essentially absent in stars
later than B5.

The Si~III~$\lambda$1417.20 photospheric line is seen in stars of spectral classes B2V--B8V, 
B6III--B8III and in all
supergiants. The maximum strength of this line is in early B supergiants. 

Si~II~$\lambda$1485.40 is seen in stars of spectral classes B3V--B8V, 
B3III--B8III, and B5I--B8I. 
It is stronger in later BVs and BIIIs. This line originates from an excited 
level. Consequently it is
purely stellar without interstellar contamination and can be used as a unique 
tracer of mid- to
late-B stars.

Si~II~$\lambda$1526.70, 1533.44 are located in a region contaminated by the 
P~Cygni profile of 
C~IV~$\lambda$1548.19, 1550.77. The resonance line of Si~II~$\lambda$1526.70 
has a narrow and deep interstellar component whereas Si~II~$\lambda$1533.44 is photospheric. 
Figure~\ref{f4} shows the change of these
lines in early B stars as a function of luminosity class. The maximum strength 
of Si~II 
$\lambda$1533.44 is reached in early B supergiants. O stars have Fe~IV lines in 
this region.

\subsection{Carbon and sulfur lines}

C~III~$\lambda$1247.37 and C~III~$\lambda$1426.45, 1427.84 are purely photospheric 
lines. 
C~III~$\lambda$1247.37 is seen in O and B stars of all luminosity
classes. It is stronger in stars later than O7. In B stars it becomes stronger 
in stars 
earlier than B2. It can be blended with Fe~IV~$\lambda$1246.8, 1247.8 and be 
affected by the N~V~$\lambda$1240 P~Cygni profile. C~III~$\lambda$1426.45, 1427.84 are seen only in 
early B stars. They 
are close to Fe~V~$\lambda$1429, 1430 in O stars.

\begin{figure}[h]
\plotone{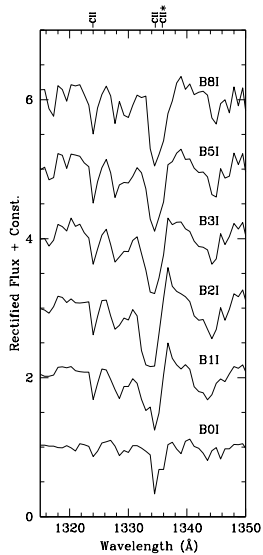}
\caption{Spectra of B supergiants showing C~II~$\lambda$1335, 1336 
(see Table 4 for a list of
stars). Excited lines (photospheric) are marked with ``$\ast$''. The spectra are
rectified.} \label{f8}
\end{figure}

C~II~$\lambda$1334.52, 1335.68 are interstellar and photospheric lines, 
respectively. Due to its
small excitation potential of 0.01 eV, the $\lambda$1335.68 line can sometimes be 
observed as an
interstellar line as well. Only the interstellar line is seen in O stars due to 
the low ionization potential of C~II. B stars, however, show either 
the two components (B0I and B0III to B3III) or a broad line formed by the two 
lines blended (B3V to B8V, B5III to
B8III and B1I to B8I). BIs show P~Cygni profiles (right panel in 
Figure~\ref{f3} and
Figure~\ref{f8}, which will be discussed later in the next section).

C~IV~$\lambda$1548.19, 1550.77 (also called C~IV~$\lambda$1550) are the strongest 
stellar lines in O stars in the satellite-UV. 
O3V to O7V and O3I to O5I show strong P~Cygni profiles that dominate the 
spectral region. In B stars these lines are strong only in 
B0V, B0III and B0I--B3I. The changes of the profiles for different luminosity 
classes are shown in  
Figure~\ref{f4}. C~IV~$\lambda$1548.19, 1550.77 are essentially absent 
in types later
than B5 when the ionization balance shifts towards C~III and C~II.

S~V~$\lambda$1501.76 is seen in most O stars of all luminosity classes and 
spectral types. In B 
stars it is seen only in early types. This line has no interstellar contribution 
since it originates
from an excited level.



\section{Population models: varying the IMF and SF}

\begin{figure}[h]
\plotone{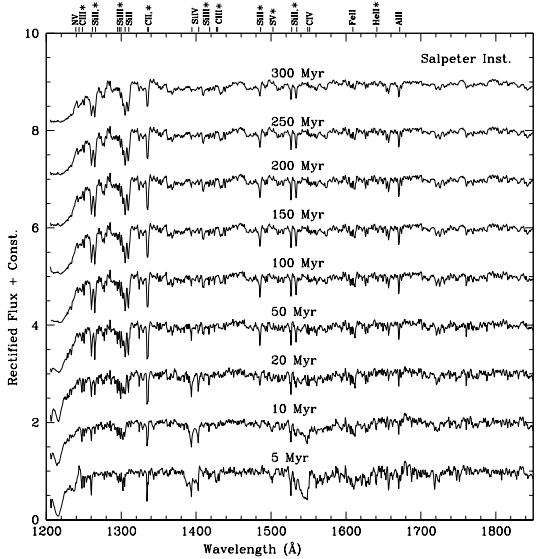}
\caption{Synthetic spectra between 5 and 300 Myr. Star-formation
law: instantaneous; IMF: Salpeter; solar metallicity. Excited lines (photospheric) are marked with ``$\ast$''. ``,$\ast$'' is used when only the
second component is an excited line. The spectra are rectified.} \label{f5}
\end{figure}

We computed a series of models to explore the effects of varying the IMF and the 
star-formation (SF) law. 
In Figure~\ref{f5} we show the results for the Salpeter IMF using an 
instantaneous SF for
 several ages (5-300 Myr). Several spectral features clearly vary with age and 
are powerful
diagnostics for age--dating 
bursts of star formation. For instance, lines such as C~III~$\lambda$1247, Si~II 
$\lambda$1261, 1265, Si~II~$\lambda$1304, 1309, C~II~$\lambda$1335, 1336, Si~IV~$\lambda$1400, Si~II 
$\lambda$1485, Si~II 
$\lambda$1527, 1533, C~IV~$\lambda$1550, and Al~II~$\lambda$1671 are quite conspicuous 
and are age dependent. 
A few resonance lines are of interstellar origin as we discussed earlier (Table 
3). One of the 
most striking spectral features is C~IV~$\lambda$1550 which 
has the strongest P~Cygni profile. This feature is typical of 
hot stars with rapidly flowing winds and becomes less pronounced 
with age. Si~IV~$\lambda$1400 shows a similar behavior. However, due to 
larger optical depth of the C$^{3+}$ ion, C~IV shows a P~Cygni profile at earlier ages, confirming the results 
by Robert et al. (1993).

\begin{figure}[h]
\plotone{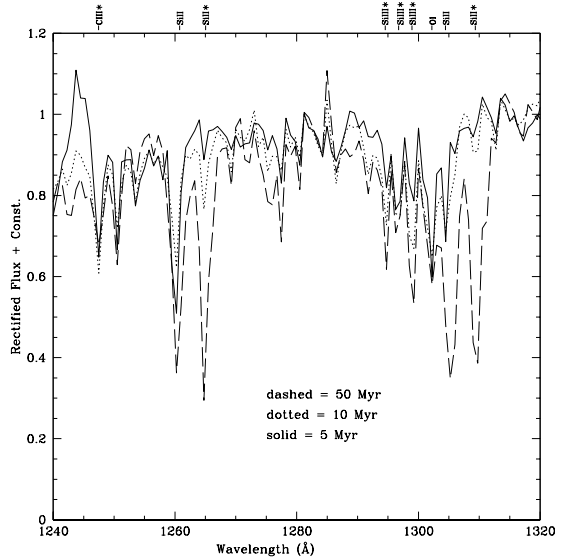}
\caption{Synthetic spectra for the wavelength region 
1240-1320~{\AA} at 5~Myr (solid line), 10~Myr (dotted line) and 
50~Myr (dashed line). Star-formation law: instantaneous; IMF: Salpeter; solar 
metallicity. Excited lines (photospheric) are marked with ``$\ast$''.
The spectra are rectified.} \label{f6}
\end{figure}

Another conspicuous feature is around $\lambda$1300. As we discussed in the 
previous section, this wavelength 
region is very rich in absorption features due to Si~III~$\lambda$1295, 1297, 
1299, O~I~$\lambda$1302, Si~II~$\lambda$1304, 1309. 
The Si~III photospheric lines become stronger at greater ages. Figure~\ref{f6} 
shows how the line profiles vary with age in this region of the
spectrum. Si~II~$\lambda$1304, 1309 are a very good age indicator; both lines get 
stronger at greater ages. Although the $\lambda$1304 line may have an interstellar contamination
in starburst spectra.

The variation of the Si~II~$\lambda$1261, 1265 doublet with age (see 
Figures~\ref{f5} and
\ref{f6}) is particularly useful to contrast stellar and interstellar lines. 
The two transitions
originate from multiplet (4) of Si~II (Moore 1950). $\lambda$1261 is from the 
ground level whereas
$\lambda$1265 comes from an excited level. In the absence of interstellar 
contamination and if wind
effects are negligible, the strength of both lines should vary with spectral 
type in a similar manner.
Si~II~$\lambda$1261 is also formed in B-star winds. Inspection of the atlas of 
Walborn et al. 
(1995b)
indicates that Si~II~$\lambda$1261 shows the same spectral-type dependence as Si~II~$\lambda$1265, suggesting
that $T_{\rm{eff}}$ is the prime driver for the behavior of both lines and that 
density effects are secondary.
Therefore the time variation of Si~II~$\lambda$1265 in Figure~\ref{f5} 
indicates the
`true' time variation of the stellar population. The Si~II~$\lambda$1261 
component carries - in
principle - the same information but the strong interstellar contribution to 
this line cannot be
separated at the resolution of Figure~\ref{f5}.
Recall that the stars used in the libraries are not free of interstellar lines; 
they
are stars in the Milky Way and have interstellar lines in their spectra. 
Interstellar lines usually are 
blended or are very close to photospheric and wind lines which makes the 
identification and removal very difficult. 
In early O stars, for instance, it is not possible to distinguish between a weak 
photospheric and interstellar 
line around Si~IV~$\lambda$1400 (Leitherer, Robert, \& Heckman 1995c). 
With the addition of the B library to our code we can investigate the relative 
proportion of the stellar and 
interstellar contribution to the resonance lines. From Figure~\ref{f6} it is 
clear that predominantly 
interstellar lines such as Si~II~$\lambda$1261 show a very small variation with 
age in contrast with
the photospheric line Si~II~$\lambda$1265. The same is true for the interstellar 
line O~I~$\lambda$1302.

\begin{figure}[h]
\plotone{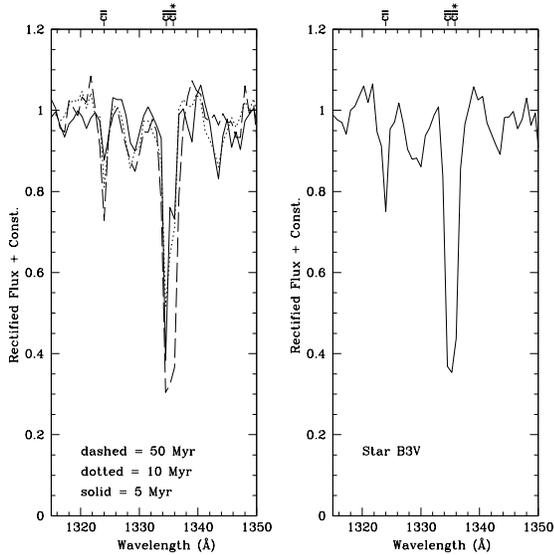}
\caption{Left panel: Synthetic C~II profiles around 1335~{\AA} at 
5~Myr (solid line), 10~Myr (dotted line) and 
50~Myr (dashed line). Star-formation law: instantaneous; IMF: Salpeter; solar 
metallicity. Right panel:
Same spectral region in a B3V star from our library (average of 4 stars - see 
Table 4 for a list of stars). Excited lines (photospheric) are marked with ``$\ast$''.
The spectra are rectified.} \label{f7}
\end{figure}

The C~II~$\lambda$1335, 1336 profile variation with age is shown in 
Figure~\ref{f7}. For small ages, 
when O stars are present, C~II~$\lambda$1335 is a deep and narrow interstellar 
line. However, for greater ages it shows as a 
broader profile. Since these spectra were generated by accessing our stellar 
libraries we searched 
for stars that have similar spectral shape. We found that stars of type B3V show 
the same broad photospheric C~II 
line (right panel in Figure~\ref{f7}). Therefore, C~II~$\lambda$1335, 1336 has a 
strong B-star contribution.
As shown in Figure~\ref{f8}, this line has a noticeable P~Cygni profile in B1 
and B2 supergiants.

\begin{figure}[h]
\plotone{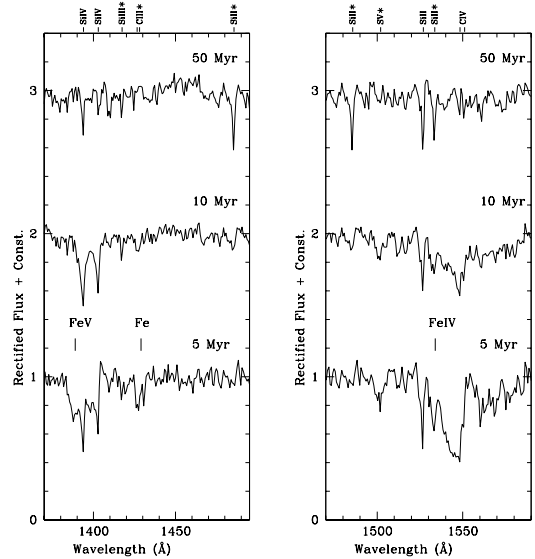}
\caption{Synthetic spectra at 5~Myr, 10~Myr, and 
50~Myr showing the spectral region around 1400~{\AA} (left) and 1550~{\AA} 
(right). The iron lines labeled as ``Fe'' are a blend of Fe~IV-V ($\lambda$1426-1429). 
Star-formation law: instantaneous; IMF: Salpeter; solar metallicity. Excited lines (photospheric) are marked with ``$\ast$''. 
The spectra are rectified.} \label{f9}
\end{figure}

The age effect in the Si~IV~$\lambda$1400 and C~IV~$\lambda$1550 profiles is shown 
in 
Figure~\ref{f9}. The P~Cygni profile typical of O stars dominates in early 
ages. It is absent
in the 50~Myr model. The pollution of P~Cygni profiles by photospheric iron 
lines has to be taken into
account when analyzing these features (Nemry et al. 1991). For instance, when O 
stars are present
(5~Myr model in Figure~\ref{f9}) we detect Fe~V~$\lambda$1388 and Fe~IV 
$\lambda$1533 but after the burst
has evolved we start detecting Si~II~$\lambda$1533 typical of B stars. 
 C~III~$\lambda$1426, 1428 also shows a strong dependence on age but there are
several iron lines (Fe~IV, V~$\lambda$1426-1429) in this region, and it is not 
possible to 
deblend their contribution accurately. Another feature that shows a strong age 
effect is 
Si~II~$\lambda$1485 which appears after 20~Myr.

\begin{figure}[h]
\plotone{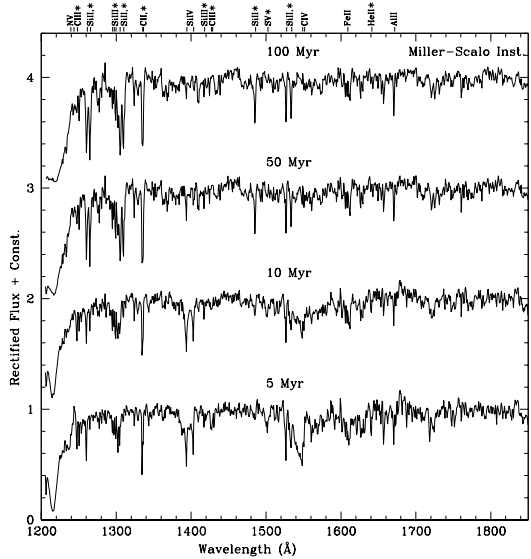}
\caption{Synthetic spectra between 5 and 100~Myr. Star-formation
law: instantaneous; IMF: Miller-Scalo; solar metallicity. The spectra are 
rectified. Excited lines (photospheric) are marked with ``$\ast$''. ``,$\ast$'' is used when only the
second component is an excited line.} \label{f10}
\end{figure}

\begin{figure}[h]
\plotone{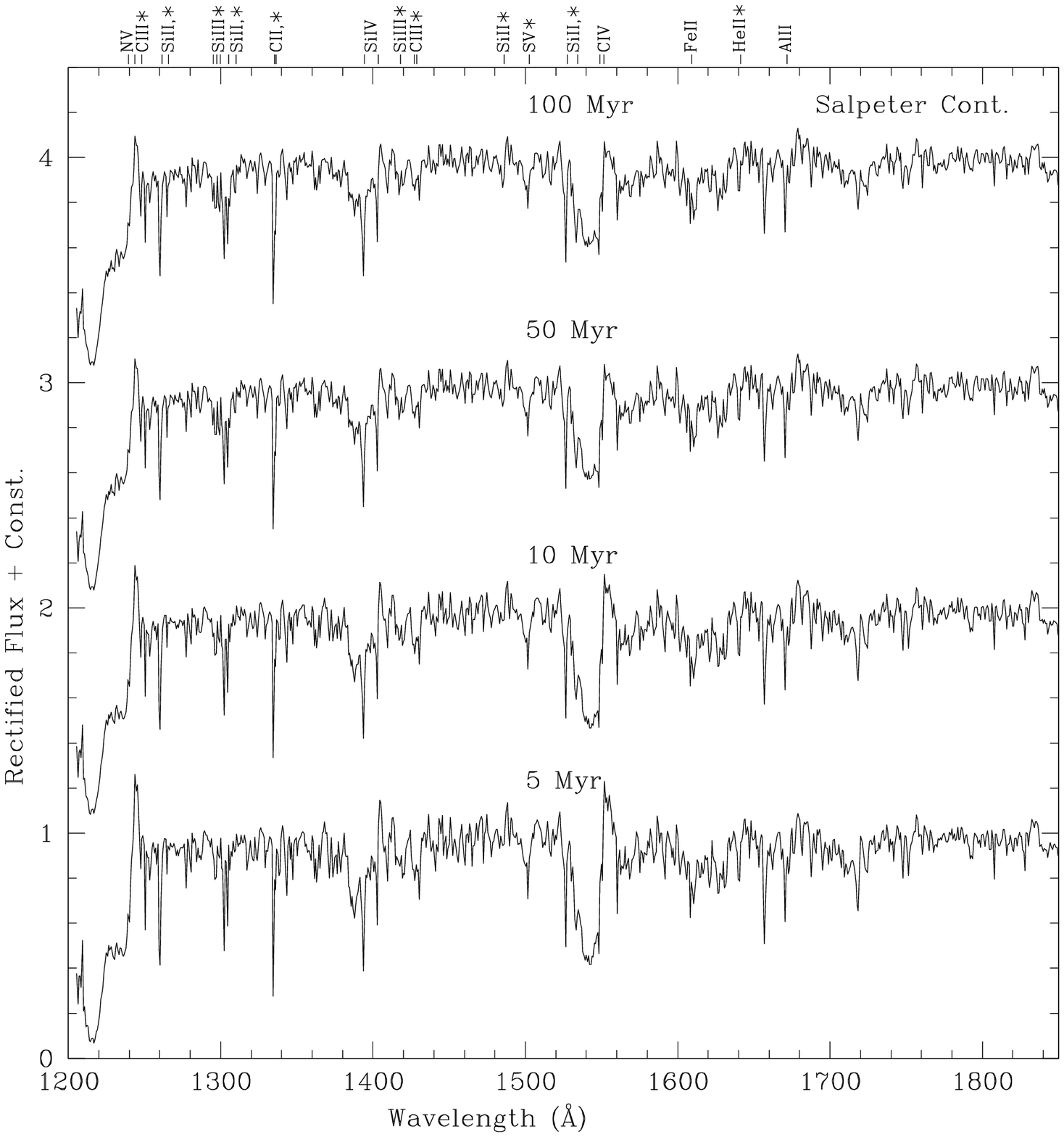}
\caption{Synthetic spectra between 5 and 100~Myr. 
Star-formation
law: continuous; IMF: Salpeter; solar metallicity. Excited lines (photospheric) are marked with ``$\ast$''. ``,$\ast$'' is used when only the
second component is an excited line. The spectra are rectified.} \label{f11}
\end{figure}

\begin{figure}[h]
\plotone{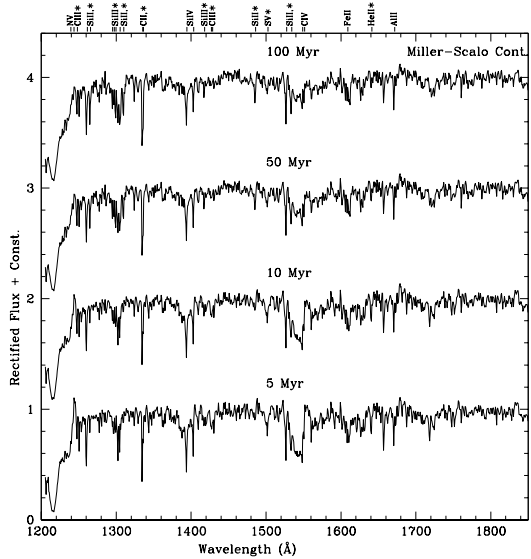}
\caption{Synthetic spectra between 5 and 100~Myr. Star-formation
law: continuous; IMF: Miller-Scalo; solar metallicity. Excited lines (photospheric) are marked with ``$\ast$''. 
``,$\ast$'' is used when only the second component is an excited line. The spectra are 
rectified.} \label{f12}
\end{figure}

Since the IMF and the star-formation history play a fundamental role in defining 
the stellar population of a star-forming region, 
we have generated models also for a Miller-Scalo IMF, which is steeper than 
Salpeter and produces
fewer massive stars. In Figure~\ref{f10} we show the UV spectra for an 
instantaneous burst
for ages 5 to 100~Myr. In comparison with a Salpeter IMF, a Miller-Scalo IMF 
produces more B and fewer O stars. Therefore
the spectra in Figure~\ref{f10} have correspondingly stronger B-star 
features at a given age than the spectra in
Figure~\ref{f5}. Compare, for instance, the strength of the features 
at $\lambda$1300 and $\lambda$1400 at 5~Myr.
In Figures \ref{f11} and \ref{f12} we 
show models for continuous SF for Salpeter and Miller-Scalo IMFs. 
The variation in the spectral shapes of strong lines such as 
Si~IV~$\lambda$1400 and C~IV~$\lambda$1550 is a powerful diagnostic
for the interpretation of the spectra of starburst galaxies. 
However, an age-IMF degeneracy is present and should be taken into account.
 For example, note the similarity between 
the model with a Salpeter IMF, continuous SF and age 100~Myr and the model with 
a Miller-Scalo 
IMF, continuous SF and age 10~Myr. This is due to the fact that both models
generate the same 
number of massive stars producing the UV light. While it is often difficult to 
disentangle age and IMF effects with UV
spectra only, the {\it duration} of the star formation can be constrained to 
some extent. Consider the Salpeter IMF models
for instantaneous (Figure~\ref{f5}) and continuous 
(Figure~\ref{f11}) star formation. The presence of
low-ionization lines from B stars and high-ionization lines from O stars is 
anticorrelated for instantaneous burst models.
The B-star features at $\lambda$1265, $\lambda$1309, or $\lambda$1335 are only 
strong when the O-star features at
$\lambda$1400, $\lambda$1501, or $\lambda$1550 have disappeared, and vice versa. 
This is no longer true in the mixed
population of Figures \ref{f11} and \ref{f12}. Both low- and 
high-ionization stellar lines are
conspicuous, indicating O and B stars. This provides a strong constraint on the 
starburst duration.

\section{Modeling NGC~1705 - a local starburst}

NGC~1705 is a blue compact dwarf at a distance of 6.2 Mpc (Meurer et al. 1995). 
A super star cluster, NGC~1705--1, probably created during a burst 10$^{7}$ years ago, 
has a core diameter of 2~pc and is responsible for 50\% of the total UV emission 
of NGC~1705 
(Melnick, Moles, \& Terlevich 1985; Meurer et al. 1992; Meurer et al. 1995). 
Because 
NGC~1705--1 is both exceptionally bright and of high surface brightness in the 
UV, it has one 
of the highest signal-to-noise UV starburst spectra taken with the HST Goddard High 
Resolution 
Spectrograph (GHRS). This galaxy was extensively discussed by York et al. 
(1990), 
Heckman \& Leitherer (1997) and Sahu \& Blades (1997), and we refer to those
papers for further information on the kinematics of its interstellar medium. An 
age of about 
10~Myr for the burst has been estimated based on the strength and shape of the 
line profiles. 
Thus, the NGC~1705--1 UV line spectrum is dominated
by the light from B stars. Furthermore, the small size of NGC 1705-1 favors 
small starburst
durations, which makes it highly likely that a single (instantaneous) population 
is
observed. Therefore this cluster is a perfect case to test our method; i.e. we 
can 
assess how accurately the new B stellar library is able to model the UV spectra 
of 
local star-forming galaxies. 

\begin{figure}[h]
\plotone{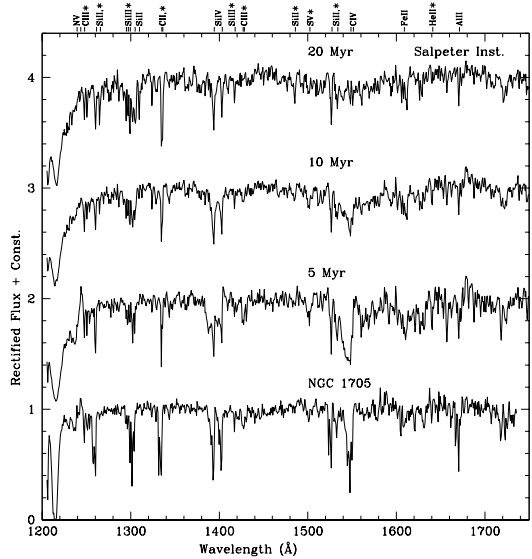}
\caption{Spectrum of NGC~1705--1 and synthetic spectra for 
ages 5, 10, and 20~Myr. Salpeter IMF, instantaneous star-formation and solar 
metallicity. Excited lines (photospheric) are marked with ``$\ast$''. ``,$\ast$'' is used when only the
second component is an excited line. The spectra are rectified.} \label{f13}
\end{figure}

NGC~1705 has an oxygen abundance of $\sim$ 0.45~Z$_{\odot}$ (Marlowe et al. 
1995), whereas the stars
in our library have solar, or somewhat subsolar metallicity. However, the 
opacity effects in the photospheric features of hot stars are significant only at 
metallicities lower than the LMC (Lejeune et al. 1997).
Therefore, for NGC~1705--1 it is adequate to use the solar metallicity tracks in 
the models
together with the stellar library.
In Figure~\ref{f13} we show the spectrum of NGC~1705--1 together with 
models for 
instantaneous bursts of 5, 10, and 20~Myr for a Salpeter IMF and solar 
metallicity. 
We have binned the spectrum of NGC~1705--1 in order to 
have the same resolution as the stellar library (0.75~{\AA}).

\begin{figure}[h]
\plotone{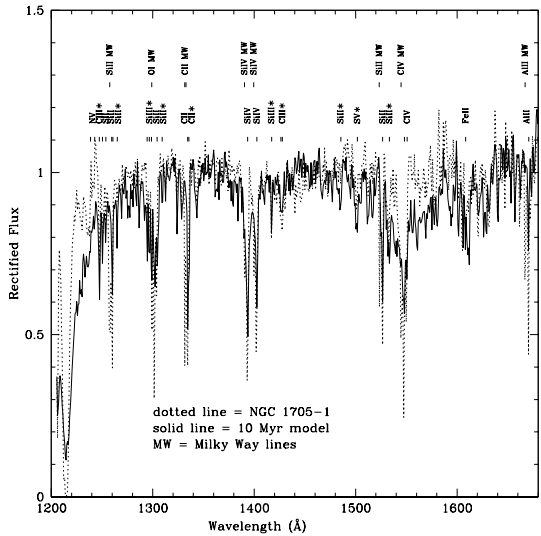}
\caption{Spectrum of NGC~1705--1 (dotted line) and synthetic 
spectrum at 
10~Myr (solid line). Salpeter IMF, instantaneous star-formation. Milky Way lines are 
identified as ``MW''. 
Excited lines (photospheric) are marked with ``$\ast$''. The spectra are 
rectified.} \label{f14}
\end{figure}

Many of the spectral features that are not well 
reproduced in our models are actually identified as interstellar lines of 
NGC~1705--1 
and/or of the Milky Way halo (see Heckman \& Leitherer 1997 for 
a detailed 
identification of these features). Figure~\ref{f14} shows the 10~Myr model 
and the spectrum 
of NGC~1705--1. 
Interstellar lines are deep and narrow and much stronger in the star cluster 
than in our models. This is
due to the fact that starburst galaxies have larger interstellar velocity 
dispersions and 
therefore their spectra present stronger interstellar lines than the library 
stars. An 
initial comparison between our models and NGC~1705--1 indicates that an age of 
5~Myr old 
is probably too small, showing too strong 
wind lines typical of O stars (N~V, Si~IV and C~IV ). The 20~Myr old model is 
probably too old, 
showing features typical of late type B stars (e.g. Si~II~$\lambda$1485) and 
lacking features 
of late O and early type B stars (e.g. S~V~$\lambda$1502) present in the star 
cluster spectrum. The best way 
to date the burst is to avoid interstellar lines and utilize 
pure stellar lines. The 
stellar lines C~III~$\lambda$1247, Si~II~$\lambda$1265, Si~III~$\lambda$1295, 1297, 
1299, 
Si~III~$\lambda$1417, C~III~$\lambda$1426, 1428, Si~II~$\lambda$1485, S~V 
$\lambda$1502, and Si~II~$\lambda$1533 
are pure photospheric lines. However, some of them are in regions very close to 
strong 
interstellar lines and should be used with caution. We chose C~III~$\lambda$1247, 
Si~III~$\lambda$1417, and S~V~$\lambda$1502 as age diagnostics. We have applied a 
$\chi$$^{2}$
test to all ages of the models for each of these lines. 
A significantly lower $\chi$$^{2}$ is found at 10~Myr suggesting that the 10~Myr 
old burst model is the one 
that best reproduces the spectral features of the star cluster, in agreement 
with Heckman \& Leitherer (1997).
 We have inspected our models in order to verify the spectral type of B stars 
which are contributing to the 
synthetic spectra. At 10~Myr there are no more O stars (the last generation of O 
stars is seen at 7 Myr), and 
most of the B-star population is composed of BVs and early BII stars. Therefore, 
the photospheric lines such as 
C~III~$\lambda$1247, Si~III~$\lambda$1417, and S~V~$\lambda$1502 seen in the 10~Myr 
model originate in these stars.

\begin{figure}[h]
\plotone{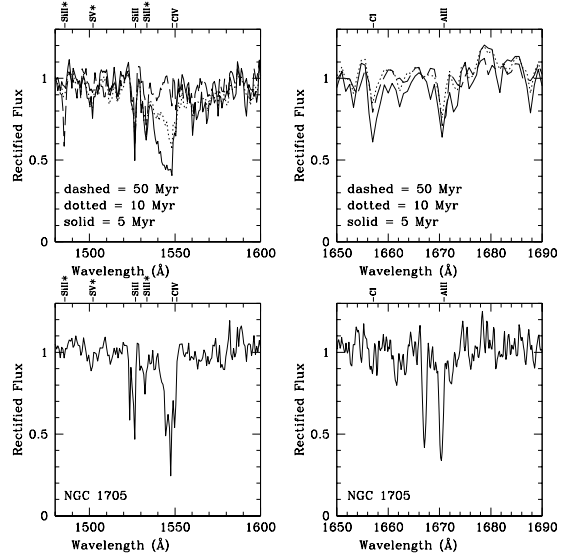}
\caption{Lower panels: Spectrum of NGC 1705--1 showing the 
spectral regions 
around $\lambda$1550~{\AA} and $\lambda$1670~{\AA}. Upper panels: Synthetic 
spectra at ages 5 (solid
line), 10 (dotted line), and
50 (dashed line) Myr. Salpeter IMF, instantaneous star-formation and solar 
metallicity. Excited lines (photospheric) are marked with ``$\ast$''. The
spectra are rectified.} \label{f15}
\end{figure}

Inspection of the observed NGC~1705--1 spectrum shows broadening and blueshifts 
of the 
resonance lines. Their origin was discussed by Heckman \& Leitherer (1997). The 
unblended resonance lines Si~II~$\lambda$1527 and Al~II~$\lambda$1671 illustrated in 
Figure~\ref{f15}
were used in their analysis. Heckman \& Leitherer concluded that these two lines 
were interstellar 
and that large-scale inhomogeneities and
macroscopic motions in the interstellar medium were the main mechanism 
generating the shape and
shifts. With the addition of the
B star library we can address the stellar contribution in more detail. 
Figure~\ref{f15} shows 5~Myr, 10~Myr and 50~Myr models for the 
Si~II~$\lambda$1527 and the Al~II~$\lambda$1671 regions. Photospheric lines 
of Si~II~$\lambda$1485 and S~V~$\lambda$1502 and wind
lines of C~IV~$\lambda$1550 clearly show an age effect while Si~II~$\lambda$1527 
and Al~II~$\lambda$1671 show no significant variation with age. This behavior suggests
that Si~II~$\lambda$1527 and Al~II~$\lambda$1671 are interstellar. For comparison, we have
marked the interstellar line CI~$\lambda$1657 (Walborn et al. 1985) in 
Figure~\ref{f15} which shows a similar behavior to Si~II~$\lambda$1527 and Al~II~$\lambda$1671.

\begin{figure}[h]
\plotone{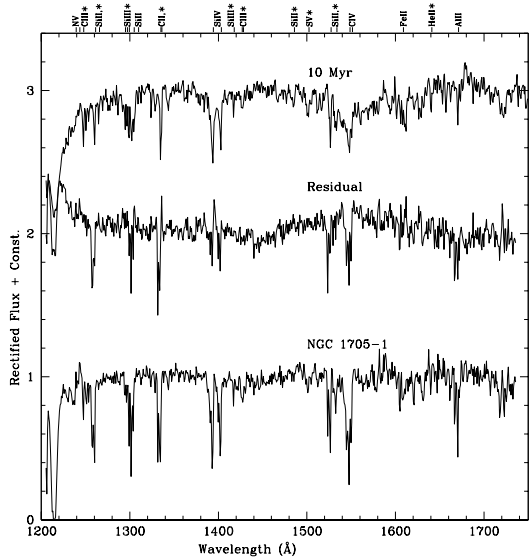}
\caption{Spectrum of NGC 1705--1, synthetic spectrum at 
10~Myr, and residual spectrum (10~Myr model subtracted from NGC 1705--1).
Excited lines (photospheric) are marked with ``$\ast$''. ``,$\ast$'' is used when only the second component is an excited line. 
The spectra are rectified.} \label{f16}
\end{figure}

A more quantitative way of identifying interstellar (IS) lines requires measurements of their full width at 
half-maximum (FWHM). Pure IS lines should have the same FWHM independent of age and any time
dependent broadening would be due to a stellar 
component. The FWHM of Al~II~$\lambda$1671 in our models remains of the order of $\sim$ 1.2~{\AA}
from 8 to 40 Myr. After that the stellar component starts dominating the profile and the
FWHM reaches $\sim$ 1.7~{\AA} at 60 Myr. Therefore, since Al~II~$\lambda$1671 is pure
interstellar at 10 Myr, we confirm the interstellar nature of Al~II~$\lambda$1671 in
NGC~1705--1. Si~II~$\lambda$1261 shows a similar behavior and it is also interstellar at
10~Myr. Other lines such as C~II~$\lambda$1335, Si~II~$\lambda$1527, Si~IV~$\lambda$1400, and C~IV~$\lambda$1551
have blended components and/or extended blue wings which affect the measurements of the FWHM. 
We estimate that Si~IV~$\lambda$1400 has about equal contributions from B-star photospheric
and from interstellar lines. In order to measure the blueshifts of these IS lines in NGC~1705--1, we have generated a 
residual spectrum by subtracting the 10 Myr model from the spectrum of NGC~1705--1. The residual spectrum 
shown in Figure~\ref{f16} contains IS lines of NGC~1705--1 and Galactic lines. The blueshift determination 
requires accurate measurements of the central wavelengths of each line. However, most of the residual 
lines have complex profiles, except for the pure IS lines of Si~II~$\lambda$1261 and Al~II~$\lambda$1671. They are blueshifted by 56~km~s$^{-1}$ and 100~km~s$^{-1}$, respectively. 
Therefore, we conclude that an average blueshift of 78~km~s$^{-1}$ is due to a directed
outflow of the IS medium in NGC~1705--1. 

\section{Modeling high-$\it{z}$ galaxies}

\begin{figure}[h]
\plotone{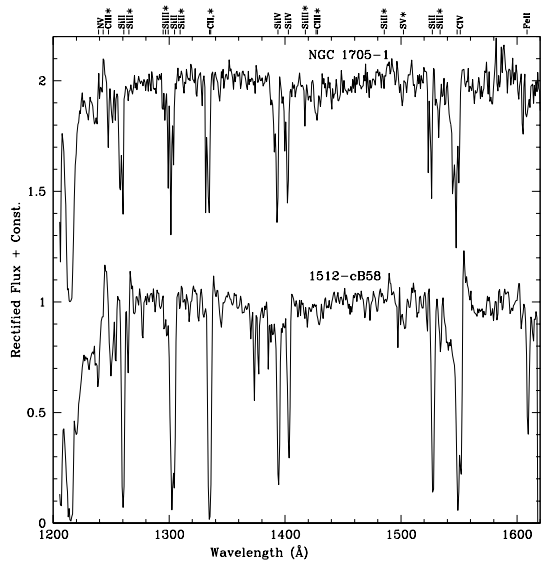}
\caption{Keck spectrum of 1512--cB58 (Pettini et al. 1997) and 
GHRS spectrum of NGC~1705--1 (Heckman \&
Leitherer 1997). Excited lines (photospheric) are marked with ``$\ast$''. ``,$\ast$'' is used when only the
second component is an excited line. The spectra are rectified and plotted in the restframe wavelength 
system.} \label{f17}
\end{figure}

We have selected the star-forming galaxy 1512--cB58 as a first
application of our models to galaxies whose stellar content is still largely 
unknown. 1512--cB58 is at {\it z}~=~2.723 and was serendipitously 
discovered by Yee et al. (1996) during the 
observation of the cluster MS1512+36 at {\it z}~=~0.373 (Gioia \& Luppino 1994) 
with the 3.6m 
Canada-France-Hawaii Telescope as part of the CNOC project (The Canadian Network 
for 
Observational Cosmology). Ellingson et al. (1996) estimated that 1512--cB58 is a 
factor of 20
brighter in the $\it{V}$ band than galaxies in the sample of Steidel et al. 
(1996). 
Seitz et al. (1998) using HST images have modeled the 
gravitational lens and showed that 1512--cB58 is lensed into a gravitational 
fold arc 
by the cluster. The part of the source of 1512--cB58 which is mapped into the arc is 
reconstructed 
by their models, and its magnification is found to be $\mu_{\rm {arc}} \ge 50$. 
They suggest that this
galaxy is not intrinsically bright but just a `normal' star-forming galaxy at 
high-$\it{z}$.
Because of its flux magnification, observations with Keck can reach a 
signal-to-noise 
of 50, better than achievable for local starburst galaxies, in just 11,400 s. We 
used the spectrum shown in 
Pettini et al. (1999), which has a resolution of 3.5~{\AA} 
(=~0.94~{\AA} in the restframe), in our analysis. The spectrum was
continuum normalized by dividing the spectrum by a spline fit to the continuum, 
following the
same procedure as applied to the stars in the library. We confirm the 
identification of the 
absorption lines by Yee et al. (1996) and used the IRAF task {\it dopcor} to 
correct for the redshift. Figure~\ref{f17} shows the spectrum of 1512--cB58 
together with the
spectrum of NGC~1705--1 plotted in rest wavelengths. 
The absorption lines around 1370~{\AA} in the spectrum of 1512--cB58 are 
identified as Mg~II $\lambda$2796, 2803 at {\it z}~=~0.82. 
Some telluric sky residuals are seen at 1500~{\AA}.

The limitations of our method and the quality of the spectrum should be taken 
into account before choosing 
the models that are suitable for 1512--cB58. Two main issues should be 
considered. 
(i) Our method works better for objects closer to solar metallicity since the 
library is composed of
stars from the solar vicinity. Therefore, for galaxies at high-{\it z}, which 
are expected to have
lower metallicity (Pettini 1999), we can only
change the metallicity of the tracks in the models while using the available 
stellar library.
Later in this section, we
will show that this change is able to produce spectral features similar to the 
ones found in 1512--cB58. 
(ii) The biggest concern in this method is the contamination of 
diagnostic lines 
by interstellar lines. Interstellar lines are the strongest lines in the 
spectrum of 1512--cB58 and,
even more than in NGC~1705--1, they contaminate all B-star wind lines. The 
strength of the 
interstellar lines is due to a large interstellar velocity dispersion 
caused by large-scale inhomogeneities and macroscopic motions in the 
interstellar medium. 

Nevertheless, there are several features that can be used to analyze the stellar 
content of
1512--cB58. 
Two of the most remarkable features in the spectrum of 1512--cB58 are the 
P~Cygni profiles of N~V~$\lambda$1240 and C~IV~$\lambda$1550. Dey et al. (1997) have identified 
P~Cygni profiles in another high-{\it z} galaxy, the radio galaxy 4C~41.17 at 
{\it z} = 3.8. They concluded that the
P~Cygni profiles are formed in stellar winds of massive stars in 4C~41.17. It is 
unlikely that a galactic outflow 
is the origin of the P~Cygni profiles due to the large outflow mass required to 
produce them. The same
reasoning can be used to infer that the P~Cygni profiles in 1512--cB58 are {\it stellar}, as 
opposed to {\it
interstellar}. Therefore, the presence of N~V~$\lambda$1240 and C~IV~$\lambda$1550 
indicates the existence of O stars in 1512--cB58. 

\begin{figure}[h]
\plotone{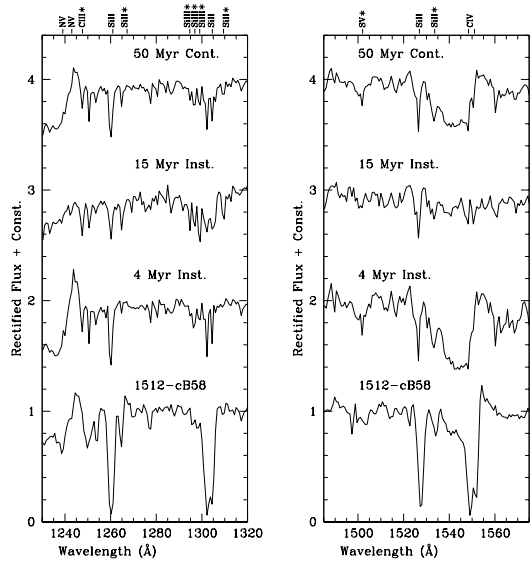}
\caption{Spectral regions around $\lambda$1280 {\AA} and 
$\lambda$1530 {\AA} of 1512--cB58. Synthetic spectrum for 
Salpeter IMF, instantaneous star-formation at 4 and 15~Myr, and continuous 
star-formation at 50
Myr; solar metallicity. Excited lines (photospheric) are marked with ``$\ast$''.
The spectra are rectified.} \label{f18}
\end{figure}

Figure \ref{f18} shows the spectral region where photospheric and wind 
lines are seen in the spectrum of 1512--cB58. 
Two models for a Salpeter IMF and instantaneous SF
at 4~Myr and 15~Myr, and one model for a Salpeter IMF and continuous SF at 
50~Myr are also shown
in Figure \ref{f18}.
In the left panel the identification of B stars in the spectrum of 1512--cB58 is 
confirmed by the 
presence of the low-ionization photospheric lines
of Si~II~$\lambda$1265 and Si~III~$\lambda$1295-1300. In the spectrum of 
1512--cB58 these lines are
more similar to the 15~Myr model, which is dominated by B stars, than to the 
4~Myr model which is dominated by O stars. 
The spectral resolution of 1512--cB58's spectrum (0.94~{\AA}) is sufficient to resolve Si~II 
$\lambda$1265 from the interstellar Si~II~$\lambda$1261. However, even though the resolution 
should be enough to separate the Si~III~$\lambda$1295-1300 lines, they are blended together with
the intrinsically very broad Si~II~$\lambda$1304 resonance line. In the right panel the 
presence of B stars is confirmed by the strong photospheric line of Si~II~$\lambda$1533. 
Thus, we conclude that the continuous SF model shows lines typical of O {\it and} B stars and is
therefore a better match to the spectrum of 1512--cB58.

We conclude that there is evidence for a composite stellar population of 
both O and B stars in the spectrum of 1512--cB58. In this case, the starburst in 
1512--cB58 is extended in time. This is consistent
with the fact that it is also spatially-extended: the entire object was placed inside the 
aperture, due to the small apparent 
size of the galaxy. This is very different from the super star cluster,
NGC~1705--1. The integrated spectrum of NGC~1705--1 comes from a single point 
source whereas in 1512--cB58 the
integrated light comes from the whole galaxy. In the spectrum of NGC~1705--1 
there are no signatures of O stars, 
like the P~Cygni profiles of N~V~$\lambda$1240 and C~IV~$\lambda$1550. On the 
other
hand, there are many features typical of B stars as we discussed in Section V. 

\begin{figure}[h]
\plotone{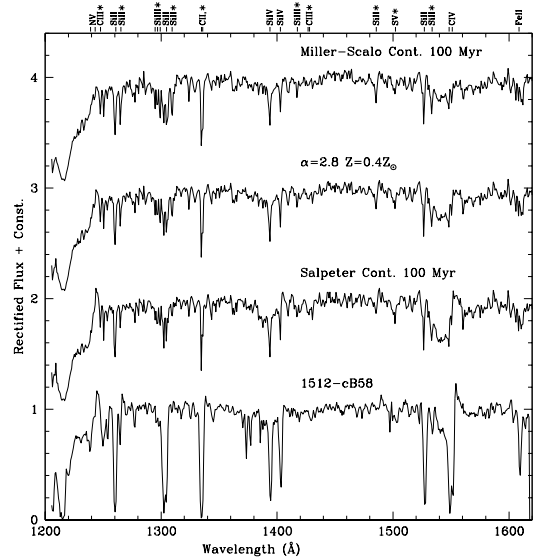}
\caption{Spectrum of 1512--cB58 and synthetic spectra for 
three models: Salpeter IMF,
continuous star-formation, 100~Myr, solar metallicity; Miller-Scalo IMF, 
continuous star-formation, 100~Myr, 
solar metallicity; IMF slope $\alpha$=2.8, continuous star-formation, 100~Myr,
metallicity $Z = 0.4$~Z$_{\odot}$. Excited lines (photospheric) are marked with ``$\ast$''. ``,$\ast$'' is used when only the
second component is an excited line. The spectra are rectified.} \label{f19}
\end{figure}

We have generated models for 1512--cB58 taking into account that spatially-extended 
starbursts  
are represented better by a continuous SF than by an instantaneous burst.
Figure~\ref{f19} shows three 
synthetic 
spectra, together with the spectrum of 1512--cB58. The first model was generated 
with a Salpeter IMF for 
continuous SF and solar metallicity. The second model was 
generated with a Miller-Scalo IMF for continuous SF and solar metallicity.
Both models are at 100~Myr. The precise age has little effect on the computed 
spectrum in continuous SF
models if {\it t}~$\ge$~20~Myr. Most of the features shown in both models are also seen 
in the spectrum of 1512--cB58. As in NGC~1705--1 the interstellar lines are 
stronger in the 
galaxy than in our models due to macroscopic motions in the ISM of both objects. Another 
major difference 
between the models and the spectrum of 1512--cB58 is the fact 
that the wind lines are too strong in the Salpeter IMF model and too weak in the 
Miller-Scalo IMF model with solar
metallicity.
Taking into account that high-{\it z} galaxies may have SMC- to LMC-like metallicity 
(Pettini
1999) and that properties of the absorption lines in local starbursts depend 
strongly 
on metallicity and IMF (Heckman et al. 1998; Leitherer 1999a; Robert et al. 
1999), we have 
generated a third model changing these parameters. We used an IMF slope 
$\alpha$~=~2.8 and 
sub-solar metallicity for the {\it tracks} ({\it Z}~=~0.4Z$_{\odot}$). Although the 
stellar library has solar metallicity, the change in
the metallicity of the stellar tracks and the steeper IMF causes a significant 
change in the wind lines. 
They are
not as strong as in the ones produced by the model with Salpeter IMF and solar 
metallicity. The P~Cygni profile of 
N~V~$\lambda$1240 resembles 
better the one in 1512--cB58 than the previous models. However, the strength of 
the absorption component
of C~IV~$\lambda$1550 is greater in 1512--cB58 than in the model. This is due to 
the fact that the interstellar line is much stronger in the galaxy than in the 
model. The deep absorption trough between 1545 and 1552~{\AA} 
(Figure~\ref{f18} right panel) is
entirely interstellar. This becomes immediately obvious from an inspection of 
C~IV~$\lambda$1550 profiles
in individual O stars (Walborn et al. 1985). Stellar winds never produce 
profiles which become black
immediately shortward of the emission peak. Rather, they gradually reach the 
minimum intensity after
several hundred km~s$^{-1}$, level off, and approach the continuum level after about 
2000 km~s$^{-1}$. This
simply results from the opacity vs. velocity relation in O-star winds. Applied 
to the spectrum of
1512--cB58, the stellar part of the absorption component of C~IV~$\lambda$1550 is 
the `bulge' seen at
$\sim 1540$~{\AA} in Figure~\ref{f18}. It is this `bulge' (and the emission 
component) relative to
the Si~II~$\lambda$1533 which gives the relative number of O vs. B stars.

\begin{figure}[h]
\plotone{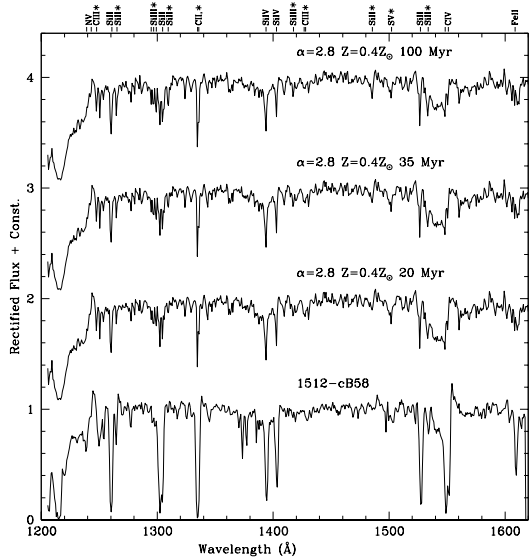}
\caption{Spectrum of 1512--cB58 and synthetic spectra for 
three models: IMF slope $\alpha$=2.8, continuous star-formation, 20, 35, and 
100~Myr, metallicity $Z = 0.4$~Z$_{\odot}$. Excited lines (photospheric) are marked with ``$\ast$''. ``,$\ast$'' is used when only the
second component is an excited line. The spectra are rectified.} \label{f20}
\end{figure}

Pettini et al. (1999) and Ellingson et al. (1996) dated the burst of 1512--cB58
at 20 and 35 Myr, respectively. We generated models for 
those ages and compared with the 100 Myr model (Figure~\ref{f20}). As mentioned 
earlier, the precise age of the burst has little effect on the
computed spectrum in continuous SF models for $\it {t}$ $>$ 20 Myr. However, the
absence of Si~II~$\lambda$1485 in the galaxy spectrum and its strength in
the models suggest that the burst is probably younger than 100 Myr but older
than 20 Myr.

We used the magnitudes measured by Ellingson et al. (1996) to calculate
the star-formation rate (SFR) for 1512-cB58 that is implied by our models
(see Table 5).
We have first increased the fluxes by 3.6 magnitudes and 1.6 magnitudes
at $V$ and $K$ respectively to correct for the effects of dust-extinction.
These values are based on the $UV$ colors of the galaxy, the
precepts of Meurer, Heckman, \& Calzetti (1999), and the empirical starburst
dust attenuation law of Calzetti (1997). We have then corrected the fluxes
downward by a factor of 50 to account for the estimated lensing magnification
(Seitz et al. 1998). Luminosities ($\lambda$$P_{\lambda}$) were computed
taking $H_0$~=~50~km~s$^{-1}$~Mpc$^{-1}$ and $q_0$~=~0.5. The resulting
intrinsic bolometric luminosity for 1512-cB58 is $\sim$ 1.5 $\times$ 10$^{12}$
L$_{\odot}$. The SFR was then
calculated by comparing these final luminosities with the predicted values
from the Starburst99 models for a Salpeter IMF, 1-100 M$_{\odot}$, $Z =$
0.4 Z$_{\odot}$, and continuous star-formation for $t$ = 100 Myr. The
resulting SFR for 1512-cB58 is estimated to be 83 and 44 M$_{\odot}$ yr$^{-1}$
based on the luminosities at rest-wavelengths of 1480 \AA\ and 5750 \AA\, respectively. 
Note that a conventional Salpeter IMF extending down to
0.1 M$_{\odot}$ would lead to SFR values that are 2.55 times larger.
For such a normal IMF, the stellar mass formed over 100 Myr would be
$\sim 2 \times 10^{10}$ M$_{\odot}$, or about 20\% of the stellar mass
of a present-day Schecter L$_*$ galaxy. We are evidently witnessing
a major galaxy-building event in 1512-cB58.
In any case, while the luminosity and implied SFR for 1512-cB58
are near the high-end of the estimated extinction-corrected values for
UV-selected `U drop-out' galaxies
at this redshift, they are not extraordinary
(e.g. Meurer, Heckman, \& Calzetti 1999).

\normalsize

{\footnotesize
\begin{table*}
\caption{1512--cB58 Data}
\footnotesize
\begin{tabular}{crrcrccccc} \hline\hline
\multicolumn{1}{c}{Band}&
\multicolumn{1}{c}{$\lambda_{\rm obs}$}& 
\multicolumn{1}{c}{$\lambda_{\rm rest}$}& 
\multicolumn{1}{c}{AB}& 
\multicolumn{1}{c}{$\lambda$F$_{\lambda}$}&
\multicolumn{1}{c}{$\lambda$$P_{\lambda}$}&
\multicolumn{1}{c}{A$_{\lambda}$}& 
\multicolumn{1}{c}{$L_{\rm final}$}& 
\multicolumn{1}{c}{$L_{\rm model}$}&
\multicolumn{1}{c}{SFR} \\
& \multicolumn{1}{c}{({\AA})} &
\multicolumn{1}{c}{({\AA})} & & (erg cm$^{-2}$ s$^{-1}$)& 
\multicolumn{1}{c}{(erg s$^{-1}$)} & &L$_{\odot}$ &L$_{\odot}$ & 
\multicolumn{1}{c}{(M$_{\odot}$yr$^{-1}$)} \\
\hline
V&5500&1480&20.64&1.1$\times$10$^{-13}$&6.15$\times$10$^{45}$&3.6&10$^{45.53}
$&10$^{43.61}$&83\\
K&21400&5750&19.61&7.30$\times$10$^{-14}$&4.08$\times$10$^{45}$&1.6&
10$^{44.55}$&10$^{42.91}$&44\\
\hline
\end{tabular}\\
\tiny
$L_{\rm final}$ = Luminosity corrected for extinction and divided by 50 (magnification factor)\\
$L_{\rm model}$ = Luminosity from Starburst99
\end{table*}

\normalsize

We have also compared our models with another high-$\it{z}$ lensed galaxy, the 
Arc 384 in the Abell 2218 
cluster (Ebbels et al. 1996), kindly provided by R. Ellis. The spectrum was 
continuum normalized
and corrected for redshift ({\it z}~=~2.515) following the same procedure as 
applied to 1512--cB58.
In Figure~\ref{f21} 
we show the spectrum of Arc 384 together with the spectrum of 1512--cB58, and 
the models with 
Salpeter and Miller-Scalo IMF described above. 
The similarities between the two lensed galaxies spectra are evident.
Unfortunately, the signal-to-noise of Arc~384 is not as high as in 1512--cB58. 
However, 
the general features are again reproduced and we can confirm that this galaxy is 
a star-forming 
galaxy at {\it z}~=~2.515.

\begin{figure}[h]
\plotone{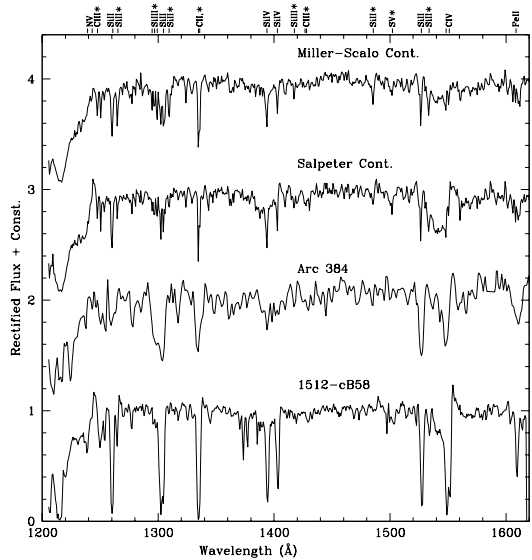}
\caption{Spectra of 1512--cB58, Arc 384, and synthetic 
spectra for two models: Salpeter IMF, continuous star-formation, 100~Myr, solar metallicity; 
Miller-Scalo IMF, 
continuous star-formation, 100~Myr, solar metallicity. Excited lines (photospheric) are marked with ``$\ast$''. ``,$\ast$'' is used when only the
second component is an excited line. The spectra are 
rectified.} \label{f21}
\end{figure}

\section{Summary and Conclusions}

The addition of the B star library to our synthesis code allows detailed studies of 
the stellar population of star-forming galaxies. We summarize our conclusions as 
follows.

(1) The contribution of B stars in the UV line spectrum becomes important for 
greater ages ({\it t}~$>$~5~Myr) when O stars have evolved. By analyzing the changes in 
the profiles 
of several UV lines such as C~III~$\lambda$1247, Si~II~$\lambda$1261, 1265,
Si~II~$\lambda$1304, 1309, C~II~$\lambda$1335, 1336, Si~IV~$\lambda$1400, Si~II 
$\lambda$1485, Si~II~$\lambda$1527, 1533, C~IV~$\lambda$1550, and Al~II~$\lambda$1671 it is possible to 
date the burst 
of star formation and to recognize the main contributors to the spectrum.

(2) Silicon is found to be a good diagnostic for spectral classification of B 
stars and lines 
such as Si~III~$\lambda$1295, 1297, 1299, Si~II~$\lambda$1265, Si~III~$\lambda$1417, 
and Si~II~$\lambda$1485 are found to be good age indicators. 

(3) Excited photospheric lines such as C~III~$\lambda$1247, Si~II~$\lambda$1265, 
Si~III~$\lambda$1295, 1297, 1299, Si~III~$\lambda$1417, C~III~$\lambda$1426,
Si~II~$\lambda$1485,
S~V~$\lambda$1502, and Si~II~$\lambda$1533 have no 
interstellar contamination and are found to be the best diagnostic of the 
stellar population. However, some are intrinsically weak and require high signal-to-noise and
high spectral resolution in order to be detected.

(4) Lines with strong P~Cygni profiles such as N~V~$\lambda$1239, 1243, Si~IV 
$\lambda$1400, and 
C~IV~$\lambda$1550 are found mainly when the stellar population is dominated by 
O stars and B
supergiants. The P~Cygni profile of C~IV~$\lambda$1550 is detected at earlier 
ages than 
Si~IV~$\lambda$1400 because of the larger optical depth of C$^{3+}$ in the wind.

(5) We used the GHRS high signal-to-noise spectrum of the super star cluster 
NGC~1705--1 
to test our models. We found that the UV line spectrum is dominated by the light 
from B stars.
The model that best reproduces the spectrum of NGC~1705--1 has a Salpeter IMF and 
instantaneous
star formation. Photospheric lines of C~III~$\lambda$1247, Si~III~$\lambda$1417, 
and  S~V~$\lambda$1502 
were used as diagnostics to date the burst of NGC~1705--1 at 10~Myr. 

(6) Broadening and blueshifts of several resonance lines are found to be stronger in 
NGC~1705--1 than in our models and are confirmed to be intrinsic of the galaxy. 
Si~II~$\lambda$1261 and Al~II $\lambda$1671 were found to be pure interstellar 
lines with an average blueshift of 78~km~s$^{-1}$ due to a directed outflow of 
the interstellar medium.

(7) We used the Keck high signal-to-noise spectrum of the gravitationally lensed 
galaxy
 1512--cB58 ({\it z}~=~2.723) as a first application of our models to high-redshift galaxies.
Direct comparison between 1512--cB58 and NGC~1705--1 showed that both spectra 
have many strong
absorption features in common. However, the P~Cygni profiles in 1512--cB58 
suggest 
that this galaxy has O stars in addition.

(8) Models with continuous star formation were found to be more adequate for 
1512--cB58 since there are
spectral features typical of a composite stellar population of O and B stars. A 
model with $Z =$~0.4~Z$_{\odot}$ 
and an IMF with $\alpha$=2.8 reproduces the stellar features of 1512--cB58 
spectrum.

We have demonstrated that our method is able to reproduce the spectral features 
of galaxies at low- and
high-{\it z}, although high signal-to-noise and high spectral resolution are  
needed in order to distinguish the stellar features typical of B stars. High 
spectral resolution is very important in this method since many of the 
photospheric lines are very close to interstellar lines and deblending becomes 
crucial.
In the two cases analyzed, NGC~1705--1 and 1512--cB58, the contribution of massive stars to the spectra is easily 
identified via diagnostic lines.
The full potential of this method can be exploited in the near future when 
high-quality spectra of local
and distant star-forming galaxies will be obtained with the new generation of 
ground-based and space
telescopes.

\acknowledgments

We are grateful to Richard Ellis for providing us with the spectrum of Arc~384, 
Ivan Hubeny and Derck Massa for helpful comments,
 Sandra Savaglio for valuable suggestions, R. Thompson for helping with IDL and echelle orders 
combination, Nolan Walborn for his comments on this manuscript, and Tommy 
Wiklind for providing the $\chi$$^{2}$ routine. We acknowledge NASA 
support for this research from an ADP grant (No. NAG5-6903) and LTSA grant
(No. NAGW-3138).




\clearpage



\end{document}